\documentclass{jfm}
\usepackage{float}
\usepackage{graphicx}
\usepackage{epstopdf, epsfig}
\usepackage{amsmath}
\usepackage{soul}
\usepackage{color}
\usepackage{siunitx}
\usepackage{subcaption}
\usepackage{commath}
\usepackage[dvipsnames]{xcolor}
\usepackage{xfrac}
\usepackage{cancel}
\usepackage{hyperref}
\newcommand{\rkbb}[1]{{\color{black}#1}}
\newcommand{\rkb}[1]{{\color{black}#1}}

\newcommand{\ub}{\textbf{u}}

\newcommand{\nb}{\textbf{n}}

\shorttitle{ Surface tension origin of the circular hydraulic jump; Experimental evidence}
\shortauthor{R.K. Bhagat, D.I Wilson, and P.F. Linden}

\title{Experimental evidence for surface tension origin of the circular hydraulic jump}

\author{Rajesh K. Bhagat$^1$
  \corresp{\email{rkb29@cam.ac.uk}},
 D. Ian Wilson$^2$, and P. F. Linden$^1$ 
 }

\affiliation{
$^1$Department of Applied Mathematics and Theoretical Physics,\\ Wilberforce Road, Cambridge CB3 0WA, UK\\
$^2$Department of Chemical Engineering and Biotechnology ,\\ Philippa Fawcett Drive, Cambridge CB3 0AS, UK}

\begin{document}

\maketitle

\begin{abstract}
{For more than a century, the consensus has been that the thin-film hydraulic jump that can be seen in kitchen sinks is created by gravity. However, we recently reported that these jumps are created by surface tension, and gravity does not play a  significant role.  In this paper, \rkbb{we present experimental data for hydraulic jump experiments conducted in a micro-gravity environment ($\approx 2\%$ of Earth's gravity) \citep{avedisian2000circular, painter2007circular, phillips2008investigation}. The existence of a hydraulic jump in micro-gravity unequivocally confirms that gravity is not the principal force causing the formation of the kitchen sink hydraulic jump.} We also present thirteen sets of experimental data conducted under terrestrial gravity reported in the literature for jumps in the steady-state for a range of liquids with different physical parameters, flow rates and experimental conditions. There is good agreement with Bhagat \textit{et al.}’s theoretical predictions. We also show that beyond a critical flow rate, $Q_C^* \propto \gamma^2 /\nu \rho^2 g$,  gravity does influence the hydraulic jumps. At lower flow rates, at the scale of the kitchen sink, surface tension is the dominating force. We discuss previously reported phenomenological and predictive models of hydraulic jumps and show that the phenomenological model – effectively a statement of continuity of radial momentum across the jump – does not allow the mechanism of the origin of the jump to be identified. However, combining the phenomenological model and  \citeauthor{bhagat2018origin}'s theory allows us to predict the height of the jump.  }

\end{abstract}

\section{Introduction}
\label{sec_introduction}

{The circular hydraulic jump, the abrupt increase in liquid depth at some distance from the impact point of a jet onto a surface that is %readily
observed when a tap is turned on in a kitchen sink, is an intriguing and practically important flow. Many cleaning and decontamination methods employ jets to wash impurities from solid surfaces, and an understanding of the flow near the point of jet impact is needed to optimise the removal of unwanted material from the surface. The simplest case to consider is that of a cylindrical liquid jet impacting normally onto a uniform planar surface. 
This configuration was sketched in the 16$^{\textrm{th}}$ century by Leonardo da Vinci and has been the subject of systematic research over the past two hundred years \citep{marusic2020leonardo}.} {Over that time }  the {consensus} view has been that {the major force responsible for the formation of} the circular hydraulic jump on the scale  observed in a kitchen sink is gravity (see figure \ref{fig:Hydro_jump} (a)).  %\citep{rayleigh1914theory,bohr2019wrong}.

Recently, we \citep{bhagat2018origin} demonstrated experimentally, supported by theory, that these jumps are caused by surface tension and gravity does not play a significant role. We conducted experiments on scales typical of those found in a kitchen sink which showed that in a thin liquid film, the circular jump produced by the normal impact of a round jet on an infinite plane is independent of the orientation of the surface. This independence of orientation with respect to the vertical demonstrated that gravity does {\it not} play a significant role in the formation of these jumps -- challenging {the accepted view. In addition, the radius $R$ of the jump predicted by our theory, in which surface tension is the dominant force, showed excellent agreement with the experiments. } %However, \cite{bohr2019wrong} contested the findings; the experimental evidence was ignored. Other researchers (cite), also supported the gravity-based theory. \cite{bhagat2019circular}, recently presented a new theoretical analysis which  showed that \cite{bhagat2018origin}'s energy based analysis is consistent with a momentum based one: it also gave the correct interfacial boundary condition.  

%{We acknowledge that for the development of any physical theory one must consider the physical reality (experiments), which is independent of the theory and ensure that the theory reflects on every element of the physical reality. Or as \cite{einstein1935can} wrote, `\textit{every element of the physical reality must have a counterpart in the physical theory}'. Among many roles which experiments play in physics, one of the main roles is to verify {or disprove} a theory or a hypothesis \citep{franklin1989neglect}. Although  \cite{bhagat2018origin} presented experimental results which gave excellent agreement with the theoretical prediction, the theory is still contested and debated. [Do we need to say this.. it's overegging things? } }

%Given the radical re-interpretation needed as a result of these findings it is not surprising that our conclusions given in
The \cite{bhagat2018origin} paper has not surprisingly been subject to scrutiny and criticism. In particular, the energy-based analysis that was used to predict the jump radius in that paper has been questioned. In response, two of us \citep{bhagat2019circular} provided a detailed analysis of the normal stress boundary condition and showed that the location of the jump was correctly predicted by conservation of momentum. 

\rkbb{Establishment of a theory relies upon experiments; we have tested the role of gravity in the formation of hydraulic jumps by changing the orientation of the surface. However, direct evidence of the role of gravity can be obtained by performing experiments in the absence of gravity or in a micro-gravity environment. Such experiments are cumbersome but have nevertheless been performed and reported by \citep{avedisian2000circular,painter2007circular, phillips2008investigation}. In this paper, we present experimental data for hydraulic jumps generated in micro-gravity ($\approx 2\%$ of Earth's gravity) environments \citep{avedisian2000circular, painter2007circular, phillips2008investigation}, and compare the results with our theory. It should be noted that the formation of a hydraulic jump in micro-gravity at approximately the same radius as in terrestrial experiments constitutes sufficient proof that gravity does not play a significant role in the formation of the jump.} 

Furthermore, to date, comparisons between \rkb{\citeauthor{bhagat2018origin}'s theory} and experiment have mostly been made with our own experimental data. If the theory is correct it should be able to explain previously published results obtained in other laboratories. Consequently, \rkb{in this paper we test} the validity of the \cite{bhagat2018origin,bhagat2019circular} theory by comparing its predictions with the experimental results for steady state jumps \rkb{for the flow configuration shown in figure \ref{fig:Hydro_jump}(a), where the liquid is allowed fall unobstructed from the edge of the plate,} reported by {other}, independent research groups. \rkb{Another purpose of this paper is to discern why previous theories concluded that kitchen sink hydraulic jumps are caused by gravity. For this, we examine previous modelling approaches -- both \textit{predictive} \citep{kurihara1946hydraulic,tani1949water,bohr1993shallow, wang2019role} and \textit{phenomenological} \citep{rayleigh1914theory, watson1964radial, bush2003influence} -- to identify their limitations and compare the predictive model with experiments.} \rkb{Subsequently, by combining phenomenological and predictive models, we obtain a result for a previously unpredicted key experimental feature, the height of the jump -- demonstrating the consistency of the \citeauthor{bhagat2018origin} theory.}

The paper is organised as follows. We briefly discuss the previous physical models and experimental results in  \S~\ref{Previous_studies2}, \rkb{which will set the backdrop for comparison between models, their validity, and most importantly the physics underpinning the formation of hydraulic jumps.} \rkb{Most of the previous studies} have stated that gravity is important: in order to determine the relative importance of gravity, we revisit the scaling relationship of \citep{bhagat2019circular} which included both gravity and surface tension in \S\ref{scaling_theory}, and identify a critical flow rate \textit{above which} gravity is important. In \S\ref{Momentum balance across the jump on a horizontal surface} we develop the theory to estimate the height of the liquid film downstream of the hydraulic jump. \rkb{In \S\ref{Revisiting scaling relationship} we derive a scaling relationship using continuity of radial momentum as the jump condition and compare it with \cite{rojas2010inertial}'s scaling relationship.} Experimental results are compared with theory in \S\ref{Results}. \rkb{ We discuss the limitations and implications of various models in \S\ref{Sec:Discussion}, and our conclusions are given in \S\ref{sec:conc}}.

\section{Previous studies}
\label{Previous_studies2}
 {As far as we are aware} the circular hydraulic jump was first documented by Leonardo da Vinci \citep{marusic2020leonardo}. The first experimental study {on hydraulic jumps was} conducted by \cite{bidone1819remou} \rkb{while \cite{belanger1828essai} gave the first theory.} {The earlier studies were concerned with jumps such as  those observed in rivers where gravity is the dominant force,}
\rkb{although in his theoretical work \cite{rayleigh1914theory} explicitly identified the kitchen sink jump as an example for his theory.}  
{Research on hydraulic jumps in radially spreading thin films began in the 1940s with models based on the thin-film boundary layer equations including gravity \citep{kurihara1946hydraulic, tani1949water}. \rkb{ \cite{watson1964radial} ignored gravity in the thin film and obtained a similarity velocity profile.}

\rkb{In these early as well as recent studies, hydraulic jumps have been described by one of the two types of models, namely (\textit{i}) predictive models, and (\textit{ii}) phenomenological models. In the former -- such as those of \cite{kurihara1946hydraulic, tani1949water, kasimov2008stationary,wang2019role, bhagat2018origin, bhagat2019circular} --  the governing equations divulge the location of hydraulic jumps by identifying the location of a singularity in velocity or film thickness gradient in the radial direction, at a finite radius. In phenomenological models -- such as those of \cite{belanger1828essai, rayleigh1914theory, watson1964radial} -- \rkb{the position of the jump is determined by equating the rate of change in momentum across the jump to the net thrust of the pressure.}

\cite{tani1949water}, following \cite{kurihara1946hydraulic}, postulated that the hydraulic jump arises due to an adverse gravitational pressure gradient which results in separation of the flow from the plane. They obtained the following result for the evolution of film height $h$ with radius $r$, 
\begin{equation}
    \frac{dh}{dr} = \frac{(5\pi \nu/Q)r^2 - h}{r -(10\pi ^2 g/3Q^2)r^3 h^3} \equiv \frac{(5\pi \nu/Q)r^2 - h}{r( 1 - \frac{1}{Fr^2})},
    \label{eq:predictive}
\end{equation}
where $Q$ is the volumetric flow rate, $\nu$ the kinematic viscosity and $g$ the acceleration due to gravity. Equation (\ref{eq:predictive}) has a singularity in $\frac{dh}{dr}$, giving a hydraulic jump, when $Fr^2 = \frac{u^2}{1.875 gh} = 1$. This model gave poor agreement with experiments, over-predicting the experimental values. \cite{watson1964radial} later showed that for kitchen sink hydraulic jumps the contribution of gravitational pressure gradient in thin film is small and can be ignored.} 

In a series of papers \citep{bohr1993shallow, bohr1996hydraulic, bohr1997averaging} investigated the circular jump produced in a thin film produced by the impact of a liquid jet  and modelled the flow using the shallow water equations. They compared experimental results obtained with water and with ethylene glycol with their model and also derived a scaling relation for $R$ based on the $Q$, $\nu$ and $g$ of the form 
\begin{eqnarray} \label{eq:RG}
R_{\textrm G} & \equiv & \frac{Q^{5/8}} {\nu^{3/8} g^{1/8}},
\end{eqnarray}
where the subscript `{G}' emphasises that this scaling includes gravity and ignores surface tension.  \rkb{Unlike \cite{kurihara1946hydraulic} and \cite{tani1949water}, they stated that $Fr =1$ -- the criterion of criticality -- cannot be used to determine the
position of the jump. To find $R$, they connected the inner and outer solutions by a shock, yielding its location.} \rkb{ However, in \cite{bhagat2018origin} we have shown -- both experimentally and theoretically --  that, for the formation of kitchen sink hydraulic jumps, the downstream film is unimportant. \rkbb{\cite{bohr1993shallow}'s model requires a smooth, radially-symmetric spreading outer film that drains down over the edge of the plate. In experiments with water on a flat plate, however, a symmetrically spreading outer film is rarely achieved, and this does not change the structure of the hydraulic jump. For example, in figure 1 of \cite{brechet1999circular}, the outer film drains from one side of the plate, while the hydraulic jump remains circular and at the same location. This example demonstrates that the flow in outer film is relatively unimportant in the formation of a kitchen sink hydraulic jump.} In \cite{bhagat2019circular}, we showed that $Fr = 1$ gives the same scaling relation, as that of  (\ref{eq:RG}), and it does not require the outer solution. Furthermore, their model gave a poor prediction of the experimental data.}  
 
}
% \diw{Should we acknowledge at some point the contribution from the Australian group [Button et al.], who were in correspondence with us?

% Also, the Indian group - Dasgupta, Tomar and Govindarajan, The European Physical Journal E, volume  38, 45 (2015) - who wrote to us?}

%\cite{watson1964radial} included the effect of viscosity in a thin film and 
\rkb{The second type of model is a phenomenological model: a model that is consistent with fundamental theory but has not been derived directly from the governing equations for the case under consideration. In this case, the condition widely used is the continuity of momentum at the jump or more specifically the phenomenological condition -- the balance of rate of loss of momentum to the thrust of the pressure at jump -- which takes the following form,  
% [I did not follow the sentence here so can't won't suggest any wording]
  \begin{equation}
    \rho \overline{u^2} h - \rho \overline{U^2} H =  \frac{1}{2}\rho g (H^2 -h^2).
    \label{eq:Phenomenological1}
\end{equation}
}
\rkb{Enforcing the phenomenological jump condition, equation (\ref{eq:Phenomenological1}) requires knowledge of the flow field on either side of the jump and \textit{the experimental measurement of the height of the jump}.   
To find the flow field, \cite{watson1964radial} solved the governing equations -- ignoring gravity -- which gave a smooth solution for the film thickness, $h(r)$ and a similarity solution for the velocity profile -- with the similarity variable $\eta = z/h(r)$ -- satisfying boundary conditions of (i) no slip on the substrate, and (ii) at the liquid/air interface $\frac{\partial u}{ \partial z} = 0$ at $z = h(r)$. The velocity profile takes the following form, 
\begin{equation}
    \ub = U(r)f(\eta) \hat{r} + U(r)\eta f(\eta) h^\prime \hat{z}.
    \label{Eq:watson_velocity}
\end{equation}

While the governing equation itself did not predict a jump, it allowed the momentum flux (LHS of  (\ref{eq:Phenomenological1})) to be estimated reasonably accurately, and, in combination with experimental measurement of height, allowed the location satisfying the jump condition, equation (\ref{eq:Phenomenological1}) , to be identified. It is instructive to notice that the wall normal direction velocity scales with $U(r)\frac{dh}{dr}$, and for an approximately flat film in the limit of $\frac{dh}{dr} \to 0$, this is negligible. However, close to the hydraulic jump where $\frac{dh}{dr} \to \infty$ it is the dominant term, which \cite{watson1964radial}'s solution failed to recognise. We will revisit this in \S\ref{scaling_theory}.}
%%%%%%%%%%%%%%%%%%%%%%%%%%%%%%%%%%%%%%%%%%%%%%%%%%%%%%%%

\begin{figure}
 % \begin{tabular}[b]{cc}
     \begin{subfigure}[h]{0.5\columnwidth}
      \includegraphics[trim=0 0 0 0,width=\linewidth]{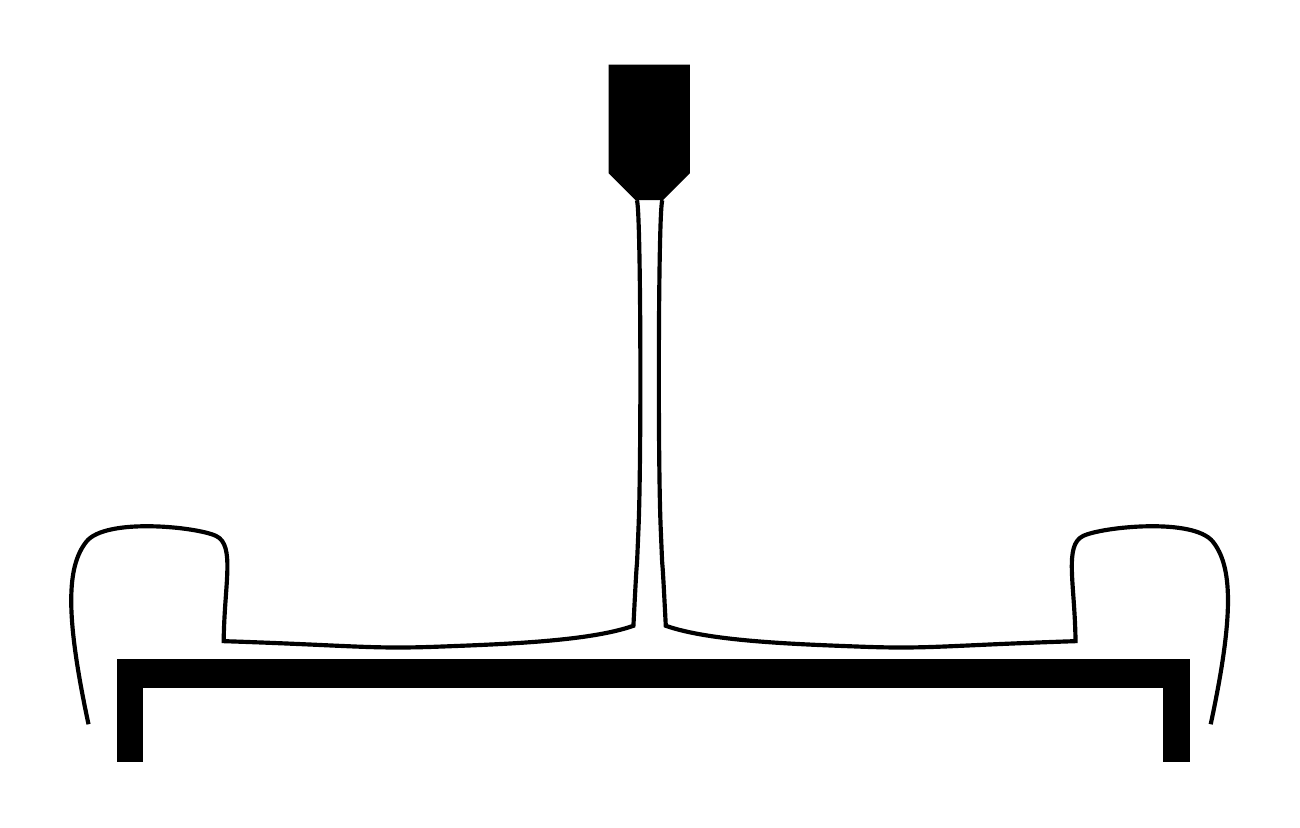}
     \caption{Hydraulic jump on a flat plate }
    %   \label{fig:2a}
    \end{subfigure}
    \hfill
   % &
      \begin{subfigure}[h]{0.5\columnwidth}
        \includegraphics[trim=0 0 0
        0,width=\linewidth]{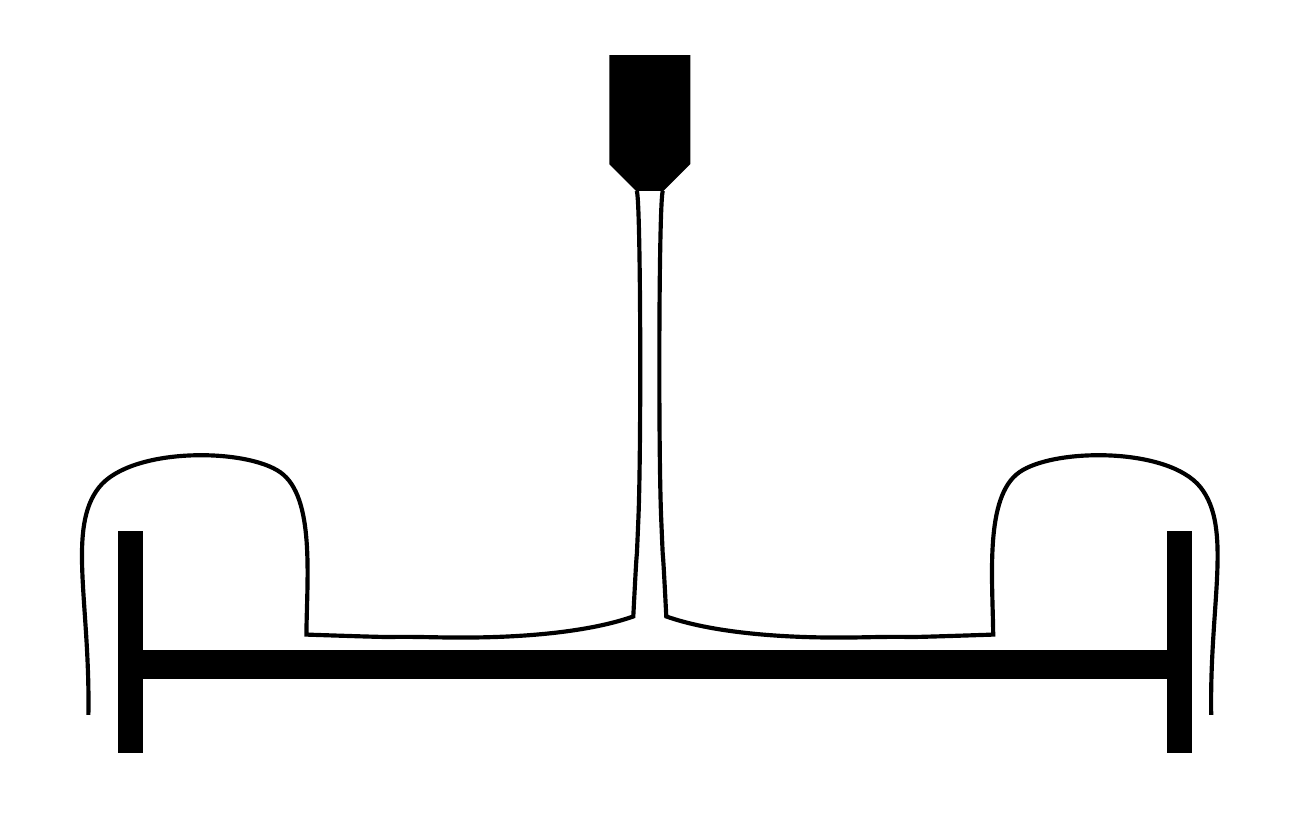}
       \caption{\rkbb{Hydraulic jump in a reservoir}}
    %   \label{fig:2b}
      \end{subfigure}
  %\end{tabular}
  
  \caption{{Hydraulic jump produced by a vertical impingement of a liquid jet on a horizontal plate (a) with no obstruction, (b) \rkbb{in a reservoir to manipulate the outer depth of the slow moving liquid film and consequently the height and the radius of the jump}. In this paper we examine the hydraulic jump produced on flat, unobstructed surfaces, \textit{e.g.} (a).}}
  \label{fig:Hydro_jump}
\end{figure}

 \rkb{\cite{bush2003influence} corrected the phenomenological model by including the influence of surface tension. They added the hoop stress term, the pressure gradient contribution arising from the interfacial tension acting at the jump, yielding  
 \begin{equation}
    \rho \overline{u^2} h - \rho \overline{U^2} H = \gamma \frac{\Delta H}{R} + \frac{1}{2}\rho g (H^2 -h^2)
    \label{eq:Phenomenological}
\end{equation}
They found that the $\gamma \frac{\Delta H}{R}$ term is only significant for smaller hydraulic jumps; for a kitchen sink scale jump $\frac{\Delta H}{R} \to 0$ and consequently, it does not have a significant effect. To verify their model, they conducted experiments in a variable depth reservoir (see figure \ref{fig:Hydro_jump}(b)). The depth of the reservoir was controlled by varying the height of the outer wall and the height of the jump was controlled by varying the depth of the reservoir. Their experiments gave good agreement with the phenomenological model, confirming that $\frac{\Delta H}{R}$ is only significant for smaller jump radii. \cite{rolley2007hydraulic} studied circular hydraulic jump formation in liquid helium and found good agreement with \cite{bush2003influence}.

In experiments where the jump radius is manipulated by varying the depth of the reservoir -- for example, increasing the depth of the reservoir increases the subcritical liquid film thickness, which consequently reduces the jump radius -- in combination with the phenomenological model, equation (\ref{eq:Phenomenological}), may suggest that the hydraulic jump \textit{originates} due to a balance between momentum and the pressure in the liquid film arising due to gravity. However, in order to examine the origin of a hydraulic jump, we consider the scenario where the liquid is allowed to flow unrestricted -- without any obstruction to modify the liquid film thickness as shown in figure \ref{fig:Hydro_jump} (a). 
\rkb{We will see that a phenomenological model is merely a relation consistent with fundamental theory, and it does not guarantee to unravel the key physics of the \textit{origin} of the jump. We will revisit the model and show its limitations and incapacity in unravelling the physics of the phenomena. }}

 \rkb{To summarise, until recently} the consensus has been that kitchen sink scale thin film hydraulic jumps are caused by gravity {and that surface tension plays at most a minor role. However, }in 2018 we \citep{bhagat2018origin} presented experimental results in which water jets impinged normally on to a horizontal wall, a vertical wall, {an inclined wall, }and {from beneath} on to a transparent ceiling, and reported that the  {radius} of the jump was independent of the orientation of the surface. These results showed unequivocally that gravity does \textit{not} play a significant role in the formation of these  jumps. {We} also presented an energy-based analysis {including both surface tension and gravity and concluded that} the jump location is caused by the surface tension of the liquid {and gravity only plays a minor role}. {Our} theory predicted the jump {radius} accurately. 

\rkb{Despite this experimental evidence, {our} theory was criticised by \cite{bohr2019wrong}, \rkb{which we address here}.   On impingement of a {cylindrical}  liquid jet {normally} onto a flat {plate} {at any orientation}, {the resulting} thin liquid film initially spread radially outwards and changed its film thickness abruptly, forming the {circular} jump, at a particular {radius $R$}. In Supplementary video 1 we show that for a particular flow configuration -- vertical impingement of a cylindrical liquid jet {normally} onto a horizontal {plate} -- beyond the jump where $r>R$, the liquid continue to spread outwards as a thicker film. Notice that the sub-critical region the liquid spreads asymmetrically however it does not affect either circular symmetry or the magnitude of jump radius any significantly -- jump remained steady, at \textit{approximately} the same radius. This implies for all practical purposed the flow inside the jump is steady, invariant and independent of sub-critical flow beyond the jump. Once the sub-critical film reaches the outer boundary, on a flat film (see figure \ref{fig:Hydro_jump} (a)); the jump location then moved inwards slightly ($\mathcal{O}$(mm)  for water), as a result of the change in the boundary condition {at the edge of the plate}. However, experiments conducted in a reservoir (such as \cite{watson1964radial}, see figure \ref{fig:Hydro_jump} (b)), the liquid film thickness increases due to the imposed boundary condition (height of the reservoir) and eventually spill over the edge of the reservoir (see figure \ref{fig:Hydro_jump} (b)). In this scenario, the imposed downstream film thickness indeed produce gravitational back pressure on the supercritical film, and as a result, the jump radius can  move inwards significantly, depending upon the height of the reservoir wall. It is instructive to note that a non-confining obstruction in sub-critical flow will not significantly affect either the circular symmetry or the jump radius. }     

% One criticism was that {we applied} a steady flow equation {to} an unsteady flow. However, {our} experiments showed that the flow field inside the hydraulic jump is independent of flow beyond the jump. On impingement of a {cylindrical}  liquid jet {normally} onto a flat {plate} {at any orientation}, {the resulting} thin liquid film initially spread radially outwards and changed its film thickness abruptly, forming the {circular} jump, at a particular {radius $R$}. Thereafter, beyond the jump where $r>R$, the liquid continued to spread outwards as a thicker film but the jump remained steady, at \textit{approximately} the same radius, {until} the film reached the outer boundary of the plate (see Supplementary video 1). The jump location then moved inwards slightly ($\mathcal{O}$(mm)  for water), as a result of the change in the boundary condition {at the edge of the plate}. {Before the liquid film reaches the edge of the plate} the jump location {and the flow upstream of the jump} \textit{is} steady, {and this state persists until} there is a change in the downstream boundary condition. 
\rkb{The governing equations of our theory apply inside the hydraulic jump, during the initial steady phase, and do not consider the flow after liquid has begun to drain from the edge of the plate when information from the edge propagates upstream against the subcritical flow downstream of the initial jump. This can cause the location of the jump to move inwards: in the limit, the jump will disappear and the jet will plunge into the pooled liquid (which is what is observed in the kitchen sink if the plug is in the drain).} 
 
 A second example, also described by our theory, is the normal impingement of a liquid jet on to a ceiling, as reported by \cite{button2010water}: a steady state exists after the jump is established, except that, at some location $r > R$, liquid falls from the ceiling due to gravity acting on the slower-moving film. We have addressed other criticisms on the consistency between momentum and energy based analyses in \cite{bhagat2019circular}. 

% The second criticism was that influence of surface tension is fully contained in the Laplace pressure, and the surface energy contribution in {our} analysis is {\textit{`wrong'}}.  \cite{bhagat2019circular} have shown that the view that the influence of surface tension is fully contained in the Laplace pressure is not correct. {From an analysis of the normal stress balance at the liquid surface we derived} the correct interfacial boundary condition {that accounts for the spatial changes in the surface area of the film as the liquid spreads radially outwards. In particular, the increase in surface area as the film thickness decreases with distance from the  point of impact of the jet is not contained in the Laplace pressure. This contribution has been neglected in previous theories that argue that surface tension is unimportant.}  

\rkb{Recently, \cite{wang2019role} presented a theoretical analysis of the hydraulic jump produced by a jet of radius $a$, and obtained the following scaled equation predicting the jump radius, 
% \diw{Are r and h dimensionless, i.e. r* = r/a and h* = h/a? If so, this needs to be confirmed as otherwise the dimensions don't add up here!}
\begin{equation}
 Re_a\left(\frac{1}{Fr_a^2} - \frac{272}{875 r^{*2}h^{*3}}\right) \frac{dh^*}{dr^*} = \frac{4}{5r^*h^{*2}}\left(\frac{69Re_a}{175}\frac{1}{r^{*2}} - \frac{3}{2h^*} \right),  
 \label{Eq:Khayat}
\end{equation}
% \begin{equation}
%  Re_a\left(\frac{1}{Fr_a^2} - \frac{272}{875 r^2h^3}\right) \frac{dh}{dr} = \frac{4}{5rh^2}\left(\frac{69Re_a}{175}\frac{1}{r^2} - \frac{3}{2h} \right),  
%  \label{Eq:Khayat}
% \end{equation}
where $r^* = r/a$, $h^* = h/a$, while the Froude and Reynolds numbers are those of the jet, $Fr_a^2 = \frac{Q}{\pi \sqrt{a^5 g}}$ and $Re_a = \frac{Q}{\pi a \nu}$. This expression (Equation 4.4 in their paper) can be rearranged and written in unscaled form
%using the constraint on mass flux, $Q = \frac{5}{4}\pi urh$ ,
for comparison with \cite{tani1949water}'s result, (\ref{eq:predictive}), giving 
\begin{equation}
    \frac{dh}{dr} = \frac{(3.86\pi \nu/Q)r^2 - h}{r( 1 - \frac{875}{272}\frac{\pi^2 g}{Q^2}r^2 h^3)} \equiv \frac{(3.86\pi \nu/Q)r^2 - h}{r( 1 - \frac{1}{Fr^2})},
    \label{eq:Khyat_predictive}
\end{equation}
where $Fr^2 = \frac{u^2}{2.06 gh}$ is the local Froude number at the location of the jump, implying $\frac{dh}{dr}$ is singular (giving a hydraulic jump) when $Fr = 1$. Comparison of (\ref{eq:predictive}) and (\ref{eq:Khyat_predictive}) indicates that they only differ in the values of the coefficients, which arises from the use of different velocity profiles:  \cite{tani1949water} used a parabolic velocity profile, while \cite{wang2019role} used a cubic one. Nevertheless, (\ref{eq:predictive}) and (\ref{eq:Khyat_predictive}) give similar solutions. The other difference between the two models is that \cite{wang2019role} incorporated the effect of gravity on the boundary layer, which for all practical purposes (and particularly for a kitchen sink scale jump) is negligible and does not change the jump radius prediction significantly. \cite{wang2019role} stated that their model is valid for high viscosity liquids; however, we did not find any physical basis for why this model should be valid for high viscosity liquids.  \cite{tani1949water} did not make any claims about viscosity, and compared its predictions with experiments conducted with water. We will revisit and discuss the model and compare it with experiments in \S~\ref{Results}.}

\cite{fernandez2019origin} reported that {our} experimental results can be explained using conventional gravity-based theory. They presented results from numerical simulations of a liquid jet flowing vertically \textit{downwards} onto a horizontal plate and compared their results with the experimental data from \cite{button2010water}, which were obtained for a jet directed {\it upwards} onto a ceiling. \rkb{Rather than supporting \cite{fernandez2019origin}'s conclusion, their result demonstrated that the jump occurs at the same location, irrespective of the orientation, confirming that gravity is unimportant. We  revisit this result in \S\ref{Sec:Discussion} and offer an explanation of how this erroneous conclusion was reached.
}

\section{\citeauthor{bhagat2018origin}'s theory}
\label{scaling_theory}

\rkb{When a round liquid jet impinges on a surface, wall friction decelerates the fluid as it spreads radially outwards. The reduction in velocity implies that the liquid film thickness should increase, while radial expansion of the liquid film implies that the film thickness should decrease, resulting in competing effects. Consequently, in the thin film, the liquid film thickness varies radially -- initially decreasing, until it reaches a minimum, and subsequently increasing until it changes abruptly, forming a hydraulic jump. Figure \ref{fig:my_label}  shows a typical liquid film thickness profile along with the velocity profiles.

In order to solve the governing equations \cite{ kurihara1946hydraulic, tani1949water, wang2019role} assumed self-similar velocity profiles, while \cite{watson1964radial} obtained the velocity profile described by equation (\ref{Eq:watson_velocity}). One of the key assumptions of these studies is that the self-similar velocity profile also satisfies the zero interfacial shear stress condition (assuming air viscosity to be negligible). For a flat film for which $\frac{dh}{dr} = 0$, $\nb = \hat{z}$, and $\hat{r}$ is the tangential direction, $(\nb \cdot \nabla)\ub = \frac{\partial u}{ \partial z} \hat{r}$, and the boundary condition applied by \cite{ kurihara1946hydraulic, tani1949water, watson1964radial, wang2019role}, namely  $\frac{\partial u}{ \partial z} = 0$  at $z = h(r)$, satisfies the zero interfacial stress condition.  

Considering the scales in figure \ref{fig:my_label}, across most of the thin film, the a flat interface can be a reasonable approximation: however, close to the jump, this approximation breaks down. Furthermore, irrespective of the velocity profile, it can be shown solely from the kinematic boundary condition that the surface velocity is
$\textbf{u}_s = u_r \hat{r} + u_r \frac{dh}{dr} \hat{z}$ \citep{bhagat2019circular}, which implies the wall normal direction velocity scales with $u_r \frac{dh}{dr}$ and close to the hydraulic jump, $u_r \frac{dh}{dr} \gg u_r$. Consequently, self-similar velocity profiles of the types used in previous studies do not represent the true velocity profile. Conversely, satisfying the zero interfacial shear stress condition implies $\frac{\partial u}{ \partial z} \neq 0$, which the previous studies do not satisfy. It is instructive to notice that, although the implicit flat film approximation prevented \cite{watson1964radial} from predicting the hydraulic jump, since the approximation breaks down only close to the jump, it allowed the momentum flux to be estimated reasonably accurately. \rkbb{In our analytical theory presented in \cite{bhagat2018origin,bhagat2019circular} we have calculated the shape factor using \cite{watson1964radial} self-similar velocity profile, as the shape factor does not change the nature of the solution, but it may give a slightly different value of the $We$ and $Fr$ numbers. However, when compared with experiments, the $We$ and $Fr$ calculated using \cite{watson1964radial} velocity profile give excellent prediction to the experimental data. }        

\cite{bhagat2019circular}'s analysis allowed the radial and wall-normal direction interfacial stresses to be calculated, which they used to show that the energy and momentum-based approaches were consistent. Here we use the \citeauthor{bhagat2018origin} theory, based on energy, to establish the relative importance of gravity and surface tension. \rkbb{\cite{askarizadeh2019role} and subsequently \cite{wang2021effects} studied hydraulic jumps using the volume of fluid (VOF) method, essentially solving the same set of equations with the same boundary conditions. \cite{wang2021effects} examined the flow over a flat plate illustrated in figure \ref{fig:Hydro_jump}(a), while \cite{askarizadeh2019role} examined the flow in a reservoir depicted in figure \ref{fig:Hydro_jump}(b). \cite{askarizadeh2019role} confirmed that until the liquid reaches the periphery of the reservoir, the hydraulic jump occurs at a radial location where $1/We + 1/Fr^2 = 1$: \cite{askarizadeh2019role} called this a developing hydraulic jump. They subsequently showed that once the liquid interacts with the outer wall at the periphery, we expect the height of the outer film to increases, and consequently the location of the jump is then determined by gravity. Although \cite{wang2021effects} did not attempt to find the condition of criticality, their study confirmed that in the absence of an outer wall or on a flat plate, the jump remained at approximately the same location, and the development of the outer film did not affect the jump significantly. These numerical studies, which confirm our experimental and theoretical observations, solved the same set of equations and boundary conditions which were presented in \cite{bhagat2019circular} (see section 4 of \cite{bhagat2019circular}). In the VOF method, the force due to surface tension is incorporated as a body force concentrated near the interfacial boundary. On a curved surface, this produces a finite radial and wall-normal direction stress, which in turn allows the zero interfacial shear stress condition, $(\nb \cdot \nabla) \ub = 0$, to be satisfied: this yields the velocity profile in the form depicted in figure \ref{fig:my_label}, and consequently gives the correct condition of criticality $1/We + 1/Fr^2 = 1$, confirming our theory and experiments.}

\begin{figure}
    \centering
    \includegraphics[width=1\linewidth]{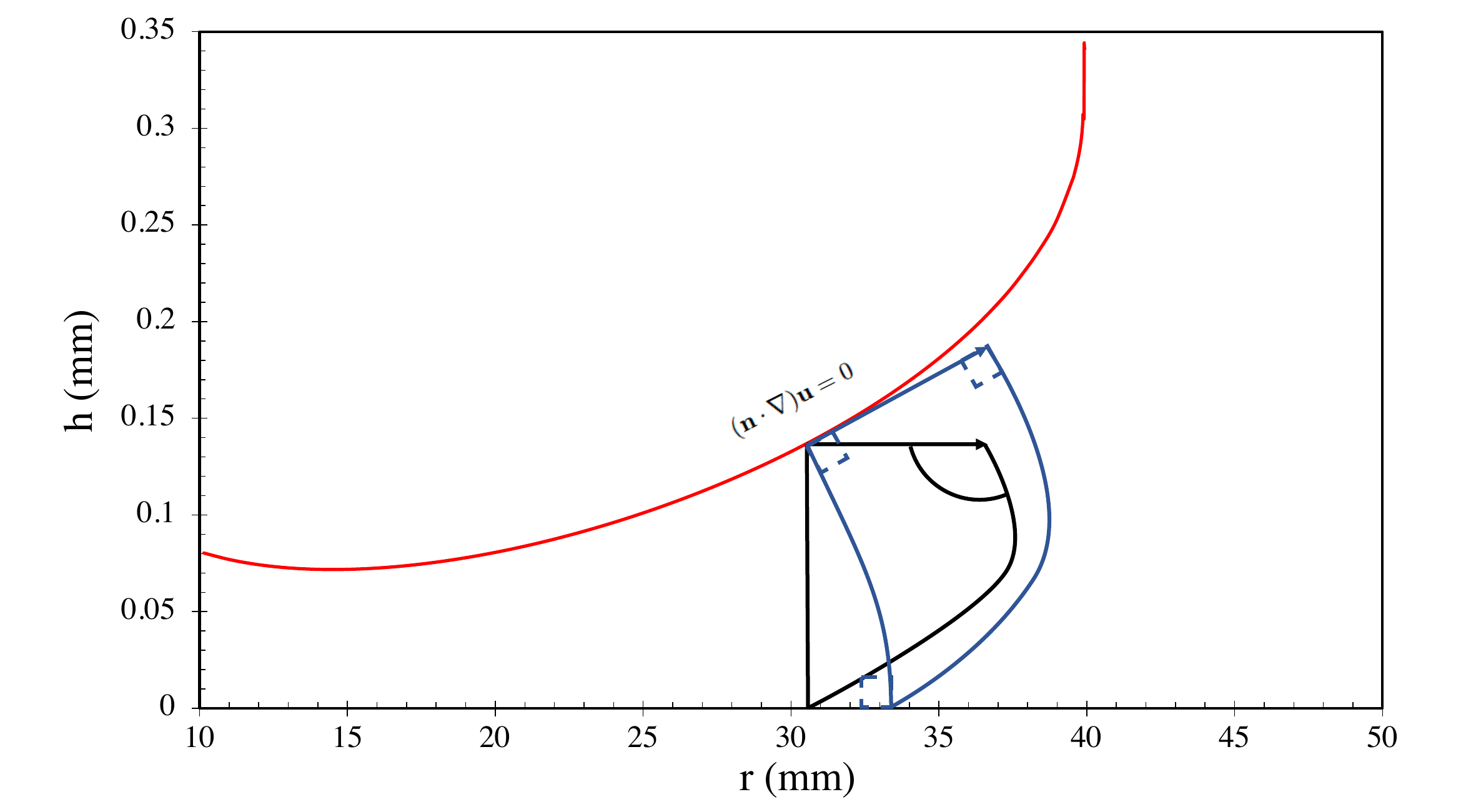}
    \caption{\rkb{Shape of the liquid film and schematic of velocity profiles for {water at 20$^{\circ}$C, $Q = \SI{2} {\litre\per\minute}$}. To satisfy the zero interfacial stress condition, a velocity profile must satisfy $(\nb \cdot \nabla)\ub = 0$ at the air/liquid interface. The blue curve shows, qualitatively, the velocity profile satisfying the zero interfacial boundary condition. The black locus shows a velocity profile in the wall-normal direction; notice that for a curved surface at the interface $\frac{\partial u}{\partial z} \neq 0$. \cite{bhagat2019circular} showed how the shape of the interface, which is dependent on viscosity and surface tension (ignoring gravity), evolve. They also showed that a similarity velocity profile which applies $\frac{\partial u}{\partial z} \neq 0$ at $z = h(r)$, violates the zero interfacial stress condition at the interface, which is particularly important near the jump where the radial velocity, $u_r$ decreases while the wall-normal direction velocity, $u_r \frac{dh}{dr}$, increases rapidly, consequently forming the jump. }}
    \label{fig:my_label}
\end{figure}
}
\subsection{Jump radius including surface tension and gravity}
\label{Jump radius including surface tension and gravity}
   {Our experiments showed that gravity is unimportant and dimensional analysis then implies that the jump radius scales as 
    % \begin{eqnarray} \label{eq:RST}
    % R_{\textrm {ST}} = 0.27 \frac{Q^{3/4} \rho^{1/4}}{ \nu^{1/4} \gamma^{1/4}},
    % \label{scaling_relation}
    % \end{eqnarray}
    
    \begin{eqnarray} \label{eq:RST}
    R_{\textrm {ST}} & \equiv & \frac{Q^{3/4} \rho^{1/4}}{ \nu^{1/4} \gamma^{1/4}},\\ \nonumber
     R & = & 0.2705 \frac{Q^{3/4} \rho^{1/4}}{ \nu^{1/4} \gamma^{1/4}}  \hspace{4mm}\textrm{\cite{bhagat2018origin}}
    \label{scaling_relation}
    \end{eqnarray}
    where $\rho$ is the liquid density, $\gamma$ is the surface tension, and the subscript `{ST}' emphasises that this scaling depends on surface tension alone \citep{bhagat2019circular}. 
    Comparison of (\ref{eq:RG}) and (\ref{eq:RST}) implies that there is a critical flowrate $Q_c \equiv \gamma^2/ \nu \rho^2 g$ above which gravity is important. We now use the results of our theoretical analysis to quantify these scaling relations. }

We extend \cite{bhagat2019circular}  {and quantify} the scaling relation including both gravity and surface tension. {We write the radial velocity as $u = u_sf(\eta)$, in terms of the surface velocity $u_s$ and the dimensionless film thickness $\eta \equiv z/h$ ($0 \leq \eta \leq 1$).}
We assume that radial flow is dominantly balanced by viscous drag, giving 
\begin{equation}
    \frac{u_s}{r} = f^\prime(0) \frac{\nu}{h^2}.
    \label{momentum}
\end{equation}
 Volume flux conservation gives
 \begin{equation}
    C_1 u_srh =\frac{Q}{2\pi},
   \label{volm_flux}
 \end{equation}
 where $C_1 = \int^1_0 f(\eta) \textrm{d}\eta$.
 Using the energy equation, including the effects of surface tension and gravity, we showed that the hydraulic jumps occurs when the flow in the liquid satisfies 

\begin{equation}
    \frac{1}{We} + \frac{1}{Fr^2} = 1,
    \label{eq:jump}
    \end{equation}
    where $We \equiv \frac{C_2 \rho u_s^2 h }{\gamma}$ is the Weber number, $Fr \equiv \frac{\sqrt{2 C_2} u_s^2}{\sqrt{gh}}$ is the Froude number and $C_2 = \int^1_0 f^2(\eta) \textrm{d} \eta$. %{Here} $u_s$,  and $h$ are the liquid velocity {at the free surface},  and the film thickness, respectively. {The shape factor }$C_2$ is \gr{the shape factor that arises due to} {determined by} the velocity profile {in the wall normal direction \citep[]{bhagat2018origin}}.
 %where $C_1$ is the shape factor arising from the velocity profile. Equation \ref{eq:1} then yields, 
%   \begin{equation}
%   \frac{\gamma}{C_2\rho u^2 h} + \frac{gh}{2C_2u^2} = 1
%   \label{jump_condn}
%   \end{equation}
  Solving  (\ref{momentum}), (\ref{volm_flux}) and  (\ref{eq:jump}) for the jump radius $R$ yields
  
 \begin{equation}
   \frac{1}{Q^*} R^{*8} + 2R^{*4} - 2 = 0,
   \label{quad}
 \end{equation}
  where the {scaled} flow rate $Q^* = \frac{2\pi C_1}{C_2 f^\prime(0)} Q_c/Q ={Q_C^*/Q}$, and $R^*  = R/ R_{\textrm {ST}} \left(\frac{C_2}{f'(0)(2\pi C_1)^3} \right)^{1/4}  $. We have used the values from Watson's similarity theory $C_1 = 0.6137, C_2 = 0.4755$ and $f'(0) = 1.402$, which give $Q_C^* = 5.78 Q_C$ and  $R^* = R/(0.2705 R_{ST})$.
 Solving (\ref{quad}) gives the predicted radius of the jump incorporating the effects of both surface tension and gravity as 
 
 \begin{equation}
     R = {0.2705R_{\textrm {ST}}\left[\sqrt{ {Q^*}^2 + 2Q^*} - Q^*\right]^{1/4}} \equiv R_{\textrm {ST}} \zeta. 
     \label{correction}
     \end{equation}

\subsection{ Jump height including surface tension and gravity}
\label{Momentum balance across the jump on a horizontal surface}

The jump marks the transition from a supercritical thin film to a subcritical thicker film, accompanied by a large decrease in liquid velocity. \rkb{We combine the phenomenological model, based on the momentum balance {at the jump}, (\ref{eq:Phenomenological}) \citep{bush2003influence}, with our theoretical model to predict the height of the jump}, giving 
  \begin{equation}
    \rho \overline{u^2} h - \rho \overline{U^2} H = \frac{\gamma (H-h)}{R} + \frac{1}{2}\rho g (H^2 -h^2),
    \label{eq:MB1}
\end{equation}
where $\overline{u}$ and $\overline{U}$ are the depth averaged velocities upstream and downstream of the jump, {respectively,}  and $H$ is the film thickness { downstream of the jump}. \rkb{We consider an ideal, sharp transitioning jump connecting upstream and downstream flows by a singularity, for which continuity {\textit{at the jump}} gives
 \begin{eqnarray}
 \label{conti1}
      \overline{u}Rh = \overline{U}RH\\ \nonumber    \Rightarrow \overline{u}h = \overline{U}H.
 \end{eqnarray}
 }
% \begin{equation} \label{conti1}
%  \overline{u}Rh = \overline{U}RH\\
%  \overline{u}h = \overline{U}H
% \end{equation} 
{From (\ref{conti1}), we can write, $\rho \bar{U^2} H = C \rho \bar{u^2} h (\frac{h}{H})$, where $C = \frac{\bar{u}^2}{\bar{u^2}} \frac{\bar{U^2}}{\bar{U}^2}$.
If we} assume that the shape factors {of the} velocity profiles across the jump {are the same,}  {implying $C= 1$,} {we find}
%   $C = {\bar{u}^2}{\bar{U^2}}/{\bar{u^2}} {\bar{U}^2}$. 
 % \begin{equation}
   % \rho \bar{u^2} h( 1 - \frac{h}{H}) = \frac{\gamma (H-h)}{R} + \frac{1}{2}\rho g H^2 (1 - \frac{h^2}{H^2})
    %\label{eq:MB1EXT}
%\end{equation}
% or,
\rkb{
\begin{eqnarray}
  \label{eq:MB1EXT1}
 \rho \overline{u^2} h \left(1-\frac{h}{H}\right) = \frac{\gamma H}{R}\left(1-\frac{h}{H}\right) + \frac{1}{2}\rho g H^2 \left(1 - {(\frac{h}{H}})^2\right)\\ \nonumber \Rightarrow
 \rho \overline{u^2} h = \frac{\gamma H}{R} + \frac{1}{2}\rho g H^2 \left(1 + \frac{h}{H}\right)
\end{eqnarray}
}
%  \begin{equation}
% \\ \nonumber \Rightarrow
%   \rho \overline{u^2} h \approx  \frac{1}{2}\rho g H^2 
%     \rho \overline{u^2} h = \frac{\gamma H}{R} + \frac{1}{2}\rho g H^2 (1 + \frac{h}{H}).
%     \label{eq:MB1EXT1}
% \end{equation}
For a \textit{surface tension dominated hydraulic jump}, $We \approx 1$, which implies $\rho \bar{u^2} h \approx \gamma$. Furthermore, for $R\gg H$, $\frac{H}{R} \to 0$ (for discussion please see \cite{bush2003influence}), and in the limit where $\frac{h}{H} \to 0$,  (\ref{eq:MB1EXT1}) can be written as, 

\begin{equation}
    \rho \bar{u^2} h \approx \gamma \approx \frac{1}{2}\rho g H^2. 
    \label{eq:MB2}
\end{equation}

Solving (\ref{eq:MB2}) yields
\begin{equation}
  H \approx \sqrt{\frac{2\gamma }{g\rho}} = \sqrt{ 2}l_c,
 \label{eq:CAP2} 
\end{equation}
where $l_c = \gamma/g\rho$ is the capillary length.

% \diw{This is the result for the surface tension dominated jump: we haven't obtained a result for the gravity dominated jump. We may need to tune our claim to reflect that we predict H for this particular case.}

%%%%%%%%%%%%%%%%%%%%%%%%%%%%%%%%%%%%%%%%%%%%%%%%%%%%%%%

\rkb{\subsection{Revisiting scaling relationship}
\label{Revisiting scaling relationship}
In \cite{bhagat2018origin}, we derived the scaling relationship, (\ref{scaling_relation}), by solving equations  (\ref{momentum}) ($ \frac{u_s}{r} = f^\prime(0) \frac{\nu}{h^2}$), (\ref{volm_flux})  ($C_1 u_srh =\frac{Q}{2\pi}$) and the predictive jump condition $We =1$ at jump. One could also use the phenomenological jump condition, given by (\ref{eq:MB2}) ($ \rho \bar{u^2} h \approx \frac{1}{2}\rho g H^2$), which yields, 
 \begin{eqnarray} \label{eq:RST1}
    R_{\textrm {PH}} & \equiv & \frac{Q^{3/4}}{ \nu^{1/4} g^{1/4} H^{1/2}}.
    \label{scaling_relation2}
    \end{eqnarray}
 \cite{rojas2013progressive} gave an identical scaling relation, which they described as being obtained in the limit of no surface tension. This scaling relationship, obtained from the phenomenological model, does not feature surface tension and thus gives the impression that the dominating mechanism for the hydraulic jump formation is gravity. 
 
%  \diw{Here we would like to draw readers attention to our experimental work \citep{bhagat2018origin}; firstly, we observed the same jump irrespective of the orientation of the surface, even when the liquid impinged on the ceiling where gravity and wall-normal direction are the same. [I don't think this needs to be stated here - focus on the model result} 
%  Secondly, 
 
 However, in \S\ref{Jump radius including surface tension and gravity} we have shown that the height of the jump, $H$, is a manifestation of the jump condition. In a conventional hydraulic jump experiment, liquid spreads radially outwards across a horizontal surface and continuity of momentum results in a liquid film beyond the jump of thickness described by equation (\ref{eq:CAP2}). On impingement at a different orientation, this relationship for the height does not hold (gravity acts in a different direction), yet the experimental evidence indicates that the jump occurs at the same location, \rkbb{and now we have included that micro-gravity experiments where the height of the outer film is irrelevant}. Furthermore, the above result does not predict the effect of changing surface tension when other parameters are kept constant: it should be noted that in these studies $H$ is a measured parameter. We have now shown that reducing the surface tension will result in a thinner downstream film and consequently a larger jump radius, while  (\ref{scaling_relation2}) will also hold.  \cite{mohajer2015circular} have verified experimentally that reducing the surface tension of water results in an expansion of the jump radius and thinner downstream film, as reported by \cite{bhagat2018origin}. The phenomenological model -- which is a momentum balance -- will still be satisfied.}

%  \diw{I think that the above section needs reworking. Focus on the modelling aspects - not the experimental evidence}

\subsection{Experimental data}
 {In this section we describe the experiments that are compared with our model and those of \rkb{\cite{bohr1993shallow} and (for more cases) of \cite{wang2019role}}, focusing on data obtained previously by independent laboratories. We restrict attention to data where the jets impacted on flat plates and the downstream thickness was not manipulated by conducting experiments in a reservoir (See figure \ref{fig:Hydro_jump}), note that a partial downstream obstacle does not change the jump radius. We also} ignore the slight adjustment ($\mathcal{O}(mm)  \textrm{for water}$) due to {the} change in boundary condition when the liquid film downstream of the jump reaches the edge of the plate and {drains off.} 
 
Table \ref{table:table1} summarises the studies considered. \cite{avedisian2000circular} (A1) examined hydraulic jumps in a drop tower facility in micro-gravity environment. \cite{painter2007circular} (PA1) and \cite{phillips2008investigation} (PH1) investigated hydraulic jumps in NASA's C-9 micro-gravity research aircraft. \cite{bohr1996hydraulic} (BO1) measured the liquid film thickness for ethylene glycol films. \cite{mohajer2015circular} (M1, M2) used water and surfactant solutions. The latter reduced the surface tension of water and increased the jump radius significantly for a given flow rate. They also measured the liquid film thickness downstream of the jump and reported that the film thickness \rkb{remains almost constant and does not vary with flow rate}. \cite{bohr1993shallow} (BO2) included experimental results for water where they varied the distance between the nozzle and the plate. \cite{brechet1999circular} (BN1) used water and varied the distance between the plate and the drain over which the falling film from the edge of the plate fell. \cite{stevens1991local} (SW1) studied heat transfer by impinging liquid jets and also measured the location of the hydraulic jump for different jet diameters. \cite{choo2016influence} (CK1) used water and varied the jet diameter. \cite{saberi2019experimental} (S1) studied the hydraulic jump on flat as well as curved surfaces. They used ethylene glycol as the working fluid for their experiments. {\cite{duchesne2014constant} (D1, D2 and D3) studied hydraulic jumps for low surface tension and high viscosity silicone oils, as well as high viscosity water-glycerine solutions (D4 and D5). \cite{hansen1997geometric} (H1, H2 and H3) studied the hydraulic jump for water and lubricating oils -- Statoil Voltway 7 and HV 46.} The physical parameters of the fluids are listed in Table \ref{table:table1}. 

% {\it I think we should include the values of the flow rates used in the table.}

 %%%%%%%%%%%%%%%%%%%%%%%%%%%%%%%%%%%%%%%%%%%%

\begin{table}
% title of Table
\vspace{-5mm}
\centering % used for centering table
\caption{Physical parameters of the fluids used in experimental studies considered.
\vspace{3mm}
}
\begin{tabular}[t]{c c c c c c c} % centered columns (4 columns)
 %inserts double horizontal lines
 \vspace*{-1mm} 
Label & Liquid & Reference &  $\rho$ & $Q_C^*$ &$ \gamma$ $\times 10^{-3}$   & $ \nu$ $\times 10^{-6}$ \\
\\

& & &($\SI{}{\kilo\gram\per\meter\cubed}$) &($\SI{}{\centi\meter\cubed\per\second}$) & ($\SI{}{\kilo\gram\per\second\squared}$) &($\SI{}{\meter\squared\per\second}$)\\[0.5ex] 
% & ($\SI{}{\milli\meter}$)&($\SI{}{\milli\meter}$)  % inserts table
%heading
%\hline % inserts single horizontal line
A1 & Water & \cite{avedisian2000circular} & 1000 & 3058 & 72 & 1.002\\
PA1 & Water & \cite{painter2007circular} & 1000 & 3058 & 72 & 1.002\\
PH1 & Water & \cite{phillips2008investigation} & 1000 & 3058 & 72 & 1.002\\
BO1 & Ethylene glycol solution & \cite{bohr1996hydraulic} & 1100 &129.7& 45 & 7.6\\
 & \cite{rojas2010inertial} &  & &  & \\
M1 & Water & \cite{mohajer2015circular} & 1000 & 3058 & 72 & 1.002\\
M2 & Water + surfactant & \cite{mohajer2015circular} & 1000 & 807.4 & 37 & 1.002\\
BO2 & Water & \cite{bohr1993shallow} & 1000 & 3058 & 72 & 1.002\\
BN1 & Water & \cite{brechet1999circular} & 1000 & 3058 & 72 & 1.002\\
SW1 & Water & \cite{stevens1991local} & 1000 & 3058 & 72 & 1.002\\
CK1 & Water & \cite{choo2016influence} & 1000 & 3058 & 72 & 1.002\\
% & 3.83 & 3.92  \\ % inserting body of the table
S1 & Ethylene glycol   & \cite{saberi2019experimental}& 1100 &91.6 & 47.5 & 12\\ 
% & 2.86 & 2.76 \\
D1 & Silicone oil - 1  & \cite{duchesne2014constant} & 950-965 &12.44& 20 & $20.4\pm 0.6$\\ 
% & 2.1 \\
% & 2.36-3.8 \\
D2 & Silicone oil - 2  &\cite{duchesne2014constant} & 950-965 &5.8& 20 & $44.9 \pm 1.5$\\
D3 & Silicone oil - 3  &\cite{duchesne2014constant} & 950-965 &2.6& 20 & $98.8\pm3$\\
D4 & Water-glycerine - 1  &\cite{duchesne2014constant} & 1190 &97.8
& 65 & $18\pm0.7$\\
D5 & Water-glycerine - 2  &\cite{duchesne2014constant} & 1220 &39.2& 65 & $44\pm1.5$\\
H1 & Water+surfactant/water & \cite{hansen1997geometric} & 1000 & 807.4 & 37 & 1.002\\
H2 & Statoil Voltway 7  &\cite{hansen1997geometric} & 875 &104& 45 & $15$\\
H3 & HV 46  &\cite{hansen1997geometric} & unknown &--& unknown & $95$\\
% & 2.1 &-  \\
% WP95/5 & \cite{bhagat2018origin} & 1000 & 474 & 42.5 & 1.274 \\
% % &2.96 &-   \\
% WP80/20 & \cite{bhagat2018origin} & 1000 & 335 & 26 & 2.30\\ 
% % &2.34 &-   \\
% SDBS & \cite{bhagat2018origin} & 1000 & 849 & 38 & 1 \\ 
% &2.78 &-   \\
%[1ex] % [1ex] adds vertical space
%\hline %inserts single line
\end{tabular}
 % is used to refer this table in the text

\label{table:table1}
\end{table}

\section{Results}
\label{Results}
%%%%%%%%%%%%%%%%%%%%%%%%%%%%%%%%%%%%%%%%%%%%%%%%%%%%%%%%%%%%%%%%%%%
\begin{figure}
 % \begin{tabular}[b]{cc}
     \begin{subfigure}[h]{0.6\columnwidth}
      \includegraphics[trim=0 0 0 0,width=\linewidth]{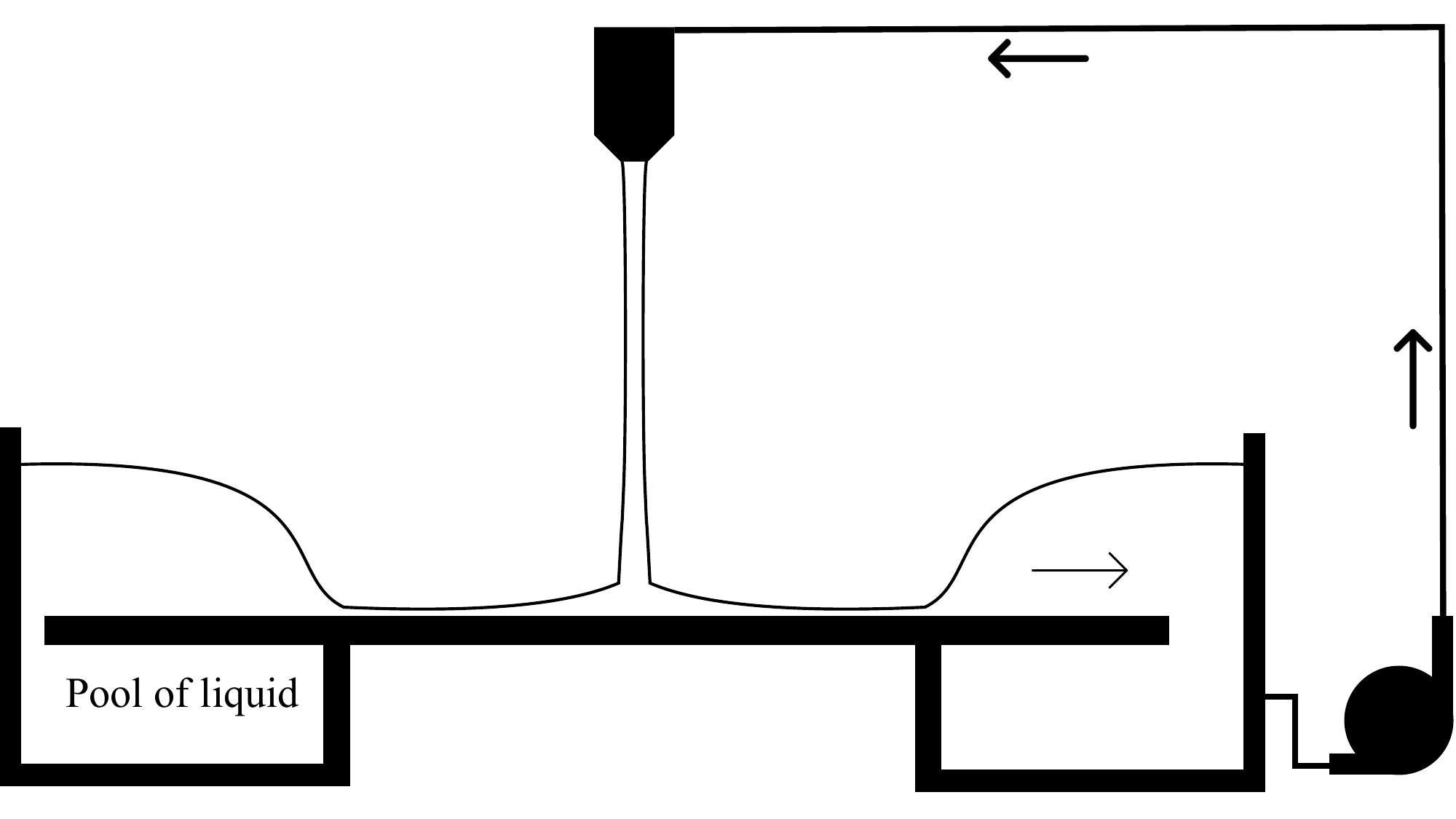}
     \caption{Hydraulic jump on a flat plate submerged in a pool of water}
    %   \label{fig:2a}
    \end{subfigure}
    \hfill
   % &
      \begin{subfigure}[h]{0.4\columnwidth}
        \includegraphics[trim=0 0 0
        0,width=\linewidth]{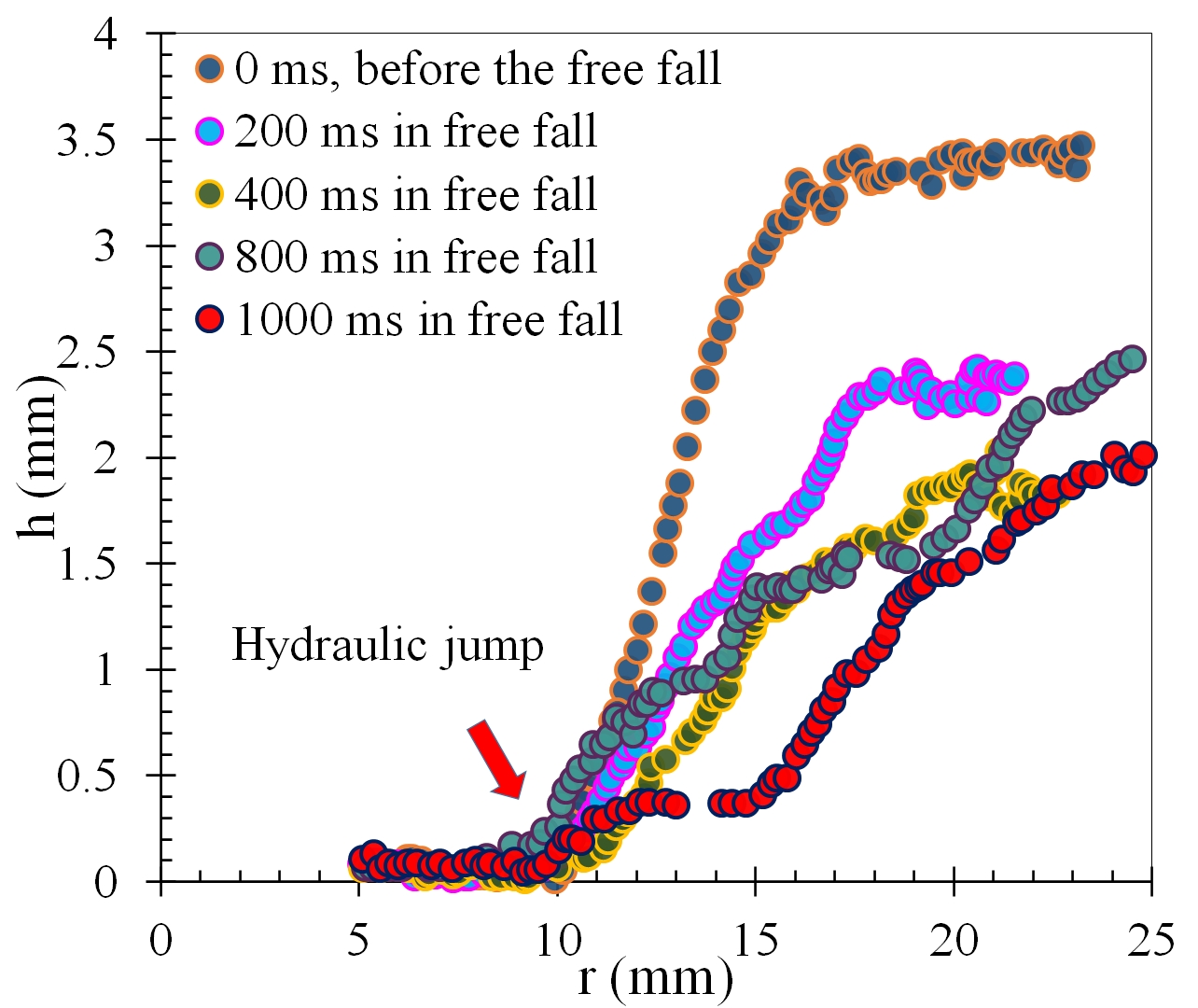}
       \caption{\rkbb{ Comparison of film thickness and jump in normal and micro gravity }}
    %   \label{fig:2b}
      \end{subfigure}
  %\end{tabular}
  
  \caption{\rkbb{(a)Hydraulic jump generated by a vertical impingement of a liquid jet on a horizontal plate submerged in a pool of liquid. (b) Comparison of liquid film thickness in gravity on earth ($g = \SI{9.8}{\meter\per\second\squared}$) and in micro-gravity (2\% of earth's gravity) after $\SI{1000}{\milli\second}$ in free fall \cite{avedisian2000circular}. The jump remains at the same location while the downstream film evolves during the free fall.}}
  \label{fig:Hydro_jump2}
\end{figure}

\begin{figure}
    \centering
    \includegraphics[width=\linewidth]{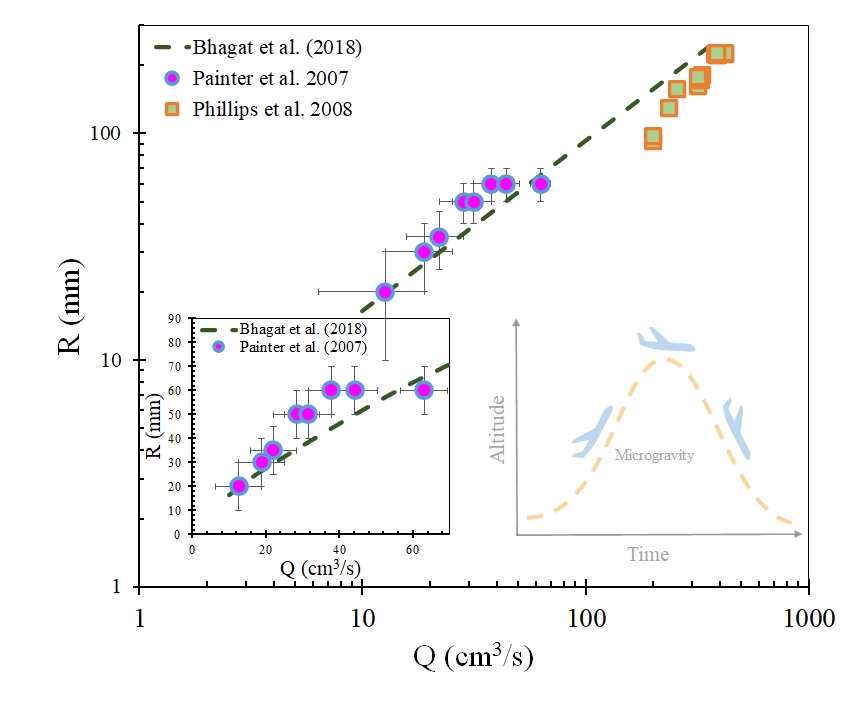}
    \caption{\rkbb{Comparison of hydraulic jump experiments conducted aboard the NASA C-9 micro-gravity research aircraft in micro-gravity environment ($\approx 2\%$ of Earth's gravity) in two separate expeditions in 2006 and 2007 by \cite{painter2007circular, phillips2008investigation}. The existence of hydraulic jump at approximately the same radius as observed in terrestrial experiments unequivocally confirms that gravity does not play a significant role in the formation of these jumps, confirming our earlier observations presented in \cite{bhagat2018origin}. The authors of \cite{painter2007circular} provided the raw data with comments, which allowed us to identify the error bars and include only reliable data points.} }
    \label{fig:zerogravity}
\end{figure}

\rkbb{\subsection{Hydraulic jumps in micro-gravity}
\label{Sec:Hydraulic jumps in micro-gravity}
We begin by considering the influence of gravity on circular hydraulic jumps. In \cite{bhagat2018origin} we changed the orientation of the surface to examine the influence of gravity on the hydraulic jump. The role of gravity in the formation of a hydraulic jump can also be studied by varying the magnitude of the gravitational constant. \cite{avedisian2000circular} carried out hydraulic jump experiments in micro-gravity in a  drop-tower facility where a reduced gravity  ($\approx 2\%$ of earth's gravity) was achieved during $\SI{1.2}{\second}$ of free fall. The experiments were carried out with the following manipulated variables: 1) gravitational acceleration (Earth's gravity and micro-gravity); 2) liquid flow rate; and 3) the outer depth of the liquid film. In order to control the outer depth, the target plate was submerged in a pool of liquid up to a predefined depth. A liquid jet impinged on to the submerged plate (see figure \ref{fig:Hydro_jump2} (a)). Based on the depth of the pool, the experiments could be classified into two further categories: i) experiments with shallow outer depth when on impingement of a liquid jet in normal gravity the hydraulic jump appears: its evolution was examined in micro-gravity, and ii) experiments carried out in a deep pool of water, when in normal gravity the hydraulic jump did not exist but a hydraulic jump appears in micro-gravity. The flow configuration employed in this study is significantly different than the (commonly used) configuration shown in figure \ref{fig:Hydro_jump}. We first highlight the differences between the flow configurations and their impact on micro-gravity testing, and subsequently compare the results with our theory. 

The flow configuration in figure \ref{fig:Hydro_jump} (a) allows the liquid film downstream of the hydraulic jump to develop freely, and consequently it does not impose any restrictions on the downstream film. However, the flow configuration employed by \cite{avedisian2000circular}, shown in figure \ref{fig:Hydro_jump2} (a), imposes a restriction on the height and the velocity of the film downstream of the hydraulic jump. We have shown that on the kitchen sink scale, the thin film upstream of the hydraulic jump remains largely unaffected by gravity, but the thicker film downstream of the hydraulic jump is influenced by gravity. Consequently, it is imperative that in micro-gravity, in response to the lack of gravitational force, the film downstream of the hydraulic jump will be affected, while the thin film upstream of the hydraulic jump will remain unaffected.

 Figure \ref{fig:Hydro_jump2} (b) compares the evolution of the hydraulic jump created by normal impingement of a liquid jet at a flow rate, $Q = \SI{6.32}{\centi\meter\cubed\per\second}$, on a pool of water of depth $\SI{4}{\milli\meter}$, in normal, and subsequently in micro-gravity during the free fall test. The experimental data demonstrate that the hydraulic jump remains at the same location while the shape of the film downstream of the hydraulic jump, which is influenced by gravity, evolves during the test. \cite{avedisian2000circular} reported that the jump radius expanded slightly as a result of micro-gravity, however, the experimental data do not support their statement (see figure \ref{fig:Hydro_jump2}). \cite{avedisian2000circular} also conducted hydraulic jump experiments in a deep pool of water, $10 \hspace{2mm}\textrm{and}\hspace{2mm} \SI{15}{\milli\meter}$ deep where, under normal gravity the hydraulic jump did not exist, but in micro-gravity, the jump appeared. For a flow rate of  $\SI{6.32}{\centi\meter\cubed\per\second}$, irrespective of the depth the hydraulic jump occurred approximately at the same location, see figures 11 of \cite{avedisian2000circular}. For a flow rate of $\SI{6.32}{\centi\meter\cubed\per\second}$ the jump radius is $\approx \SI{11.78}{\milli\meter}$ whilst the \cite{bhagat2018origin} prediction is $\SI{11.68}{\milli\meter}$.   
 
 Although \cite{avedisian2000circular} did not provide any theoretical explanation, they postulated that surface tension could be responsible for the origin of the jump. Their observations of hydraulic jumps in micro-gravity, at a finite radius, unequivocally refute that gravity is the principal force responsible for the formation of kitchen sink scale hydraulic jumps. 
 
%  In order to understand the qualitative and quantitative results presented in \cite{avedisian2000circular} we first examine the flow configuration and its implications on hydraulic jumps and subsequently we will compare the results with our theory.     

% The hydraulic jump experiments conducted in a liquid pool from where 
% In the first case scenario -- when outer liquid film was shallower -- they found that in micro-gravity  the jump radii expanded. A closer examination  In terrestrial scenario the hydraulic jump is sharp transition while in the jump was a gradually increasing ramp (see figure 9 of \cite{avedisian2000circular}). In the second case scenario, when the liquid jet was impinged in the deep pool of water (10 and 15 $\SI{}{\milli\meter}$), the hydraulic jump appeared during the free-fall. In figure 11 of their paper \cite{avedisian2000circular} show time series images of hydraulic jumps, and irrespective of the outer depth, hydraulic jump appears approximately at the same radius.  

Although \cite{avedisian2000circular} provided conclusive evidence that gravity is not important in the formation of a kitchen sink scale hydraulic jump, the scope of their study was limited by the short free fall times. \cite{painter2007circular} and \cite{phillips2008investigation} conducted hydraulic jump experiments aboard the NASA C-9 micro-gravity research aircraft by impinging liquid jets on a flat plate using the flow configuration shown in figure \ref{fig:Hydro_jump}(a), where the outer film was allowed to develop freely. The parabolic flight trajectory allowed a micro-gravity level of $\approx 2\%$ of Earth's gravity to be achieved for a considerably longer duration, of $\SI{25}{\second}$, than the drop tower experiments of \cite{avedisian2000circular}. \cite{painter2007circular} examined hydraulic jumps for flow rates ranging between $12.6 - 63  \SI{}{ \centi\meter\cubed\per\second}$. They reported the qualitative observation that on entering and exiting the micro-gravity, the jump radius fluctuated slightly but rapidly returned to its initial position, implying that at steady state in Earth's gravity or micro-gravity the hydraulic jump appeared at the same location. \cite{phillips2008investigation} carried out further experiments at higher flow rates. Figure \ref{fig:zerogravity} compares the experimental data reported in these studied in micro-gravity with our theory. The theory gives good predictions to the experimental data. \cite{painter2007circular} provided the raw data with comments which allowed us the estimate the errors associated with the data. Our theory slightly underpredicts the high flow rate experiments of \cite{phillips2008investigation}. At higher flow rates, they also reported a slight expansion of the jump radius.

\cite{wassenberg2019influence} examined the water splatter generated by impingement of a liquid jet on to a surface and concluded that water splatter increases with an increase in the jet Reynolds number and distance between the jet and the plate. For very high flow rate liquid jets, the hydraulic jump occured at a smaller radius than the predicted jump radius, which was attributed to the loss of liquid due to splatter. \cite{phillips2008investigation} did not report splattering; however, at the high flow rates studied, ranging between $15-\SI{25}{\litre\per\minute}$ significant splatter is expected and will lead to a slight underprediction. This body of work provides further evidence that gravity is not the principal force involved in the formation of kitchen sink scale hydraulic jumps.
}

 \subsection{Influence of gravity {on the} jump {radius} }
 \label{Influence of gravity in circular hydraulic jump}
 
%  {Comparing Table 1 and Table 2, Table 1 Duchesne et al. (2014) has three silicone oils, labelled 1-3, and Table 2  has two silicon oils, labelled 1-2, with different Qc* values. Are the Table 2 liquids different to silicone oils 1 and 2? If so, they should be labelled silicone oils 4 and 5 to avoid confusion}
 
\rkbb{In \S~\ref{Sec:Hydraulic jumps in micro-gravity}, we examined the hydraulic jump experiments carried out in micro-gravity, and we now consider the circular hydraulic jumps studies performed under terrestrial gravity and compare them with our theory.} In Table \ref{table:table1}, we provide the values of {the critical flow rate}, $Q_C^*$ {above which gravity becomes important }for the liquids considered. {Since $Q_C^* \propto \gamma^2/\nu \rho g$, lowering the surface tension and/or increasing the viscosity reduces $Q_C^*$.}  For water at room temperature $Q_C^* = \SI{3060}{\centi\meter\cubed\per\second} \equiv \SI{183}{\liter\per\minute}$, and for an aqeuous surfactant solution with surface tension roughly half that of water, $Q_C^* = \SI{807}{\centi\meter\cubed\per\second} \equiv \SI{48}{\liter\per\minute}$, {both of} which {are} much larger than normal kitchen sink flow rates and those in the laboratory experiments. 

{However, for silicone oils {D1, D2 and D3} the $Q_C^*$ values are 13.3, 5.8, and $\SI{2.6}{\centi\meter\cubed\per\second}$, respectively, {whereas the flow rates considered, 4.3-62.5, 6.7-40.6, 8.1-43 $\SI{}{\centi\meter\cubed\per\second}$, respectively, are mostly greater than $Q_C^*$}.} \rkb{Consequently, we expect gravity to play a role in these cases. In gravity dominated regimes we expect a hydraulic jump to occur due to adverse gravitational pressure gradient, when $Fr \approx 1$, and in surface tension dominated regimes when $We =1$. Consequently, in gravity dominated regimes the \cite{tani1949water} and \cite{wang2019role} model should give a reasonable prediction. }

\begin{figure}
 % \begin{tabular}[b]{cc}
     \begin{subfigure}[h]{0.5\columnwidth}
      \includegraphics[trim=0 0 0 0,width=\linewidth]{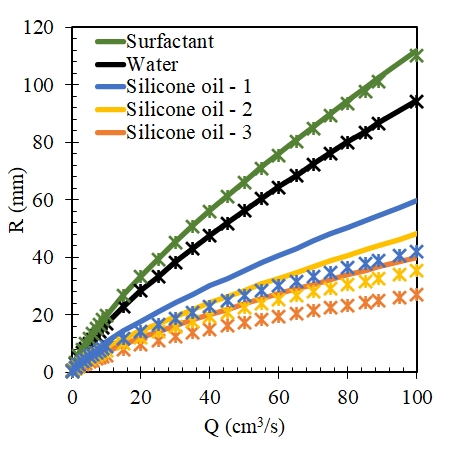}
     % \caption{Jump radius correction due to gravity}
    %   \label{fig:2a}
    \end{subfigure}
    \hfill
   % &
      \begin{subfigure}[h]{0.5\columnwidth}
        \includegraphics[trim=0 0 0
        0,width=\linewidth]{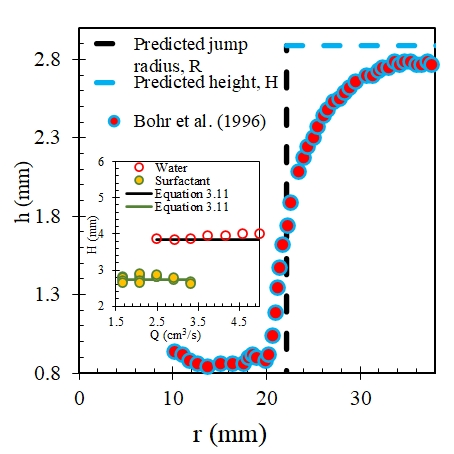}
      %  \caption{Liquid film thickness \citep{bohr1996hydraulic}}
    %   \label{fig:2b}
      \end{subfigure}
  %\end{tabular}
  \vspace{-5mm}
  \caption{(a) Comparison of {the theoretical }predictions {of the } hydraulic jump radius: {surface tension only (\ref{scaling_relation}), solid lines, and surface tension plus gravity (\ref{correction}), symbols.} %when we  consider only surface tension ( \ref{scaling_relation}) and when we consider both surface tension and gravity (\ref{eq:CAP2}), for 
  {The cases include} water, water surfactant solution, and silicone oils - 1, 2, and 3 (see Table \ref{table:table1}). For the range of flow rates shown here, {there is no appreciable difference between the two predictions for water and surfactant solutions, implying that gravity is unimportant as expected since $Q \ll Q^*_c$}. %for water and the surfactant solution the prediction from (\ref{scaling_relation}) (continous lines) and \ref{correction} (symbols) superimpose on each other. 
  {On the other hand, for the more viscous silicone oils, $Q < Q^*_c$, and the inclusion of gravity significantly reduces the predicted jump radius. }(b) Profile of surface {height} reported by \cite{bohr1996hydraulic} {(BO1)} for a jet of ethylene glycol solution falling vertically onto a horizontal plate with no weir. %Conditions: $Q = \SI{27}{\centi\meter\cubed\per\second}$, $\rho =  \SI{1100}{\kilo\gram\per\meter\cubed}$, $\nu = 7.6 \times 10^{-6} \SI{}{\meter\squared\per\second}$, $\gamma = \SI{45}{\milli\newton\per\meter}$ \citep{rojas2010inertial}. 
  Hydraulic jump radius $(R)$ and liquid film thickness at jump $H$ predicted by  (\ref{scaling_relation}) and (\ref{correction}) are shown by the black and blue dashed lines, respectively. The inset image compares the predicted and measured heights of the film, $H$, for water and surfactant solution at different flow rates.    }
  \label{fig:1}
\end{figure}

To evaluate the influence of gravity, Figure \ref{fig:1}(a) compares the {\it predicted} hydraulic jump radius considering only surface tension,  (\ref{scaling_relation}), and considering both surface tension and gravity, (\ref{correction}). For water and the aqueous surfactant solution the values are effectively identical, indicating that for these flow rates gravity does not play a significant role. However, the experimental range employed in the experiments with silicone oils (see \cite{duchesne2014constant}), gravity does play a role, {since $Q > Q^*_C$}, {and the effect of gravity is to reduce the jump radius compared with the value predicted by surface tension alone}.

 \begin{table}
% title of Table
%\vspace{-5mm}
\centering % used for centering table
\caption{Measured, $H_m$, and predicted, $H$, heights of the subcritical liquid film at the jump. \\ }
\begin{tabular}[t]{c c c c c c} % centered columns (4 columns)
 %inserts double horizontal lines
 \vspace*{-1mm} 
Liquid & Reference &  $Q$  & $Q_C^*$ & $H_m$ $\times 10^{-3}$ & $H$  $\times 10^{-3}$ \\
\\
% & & $\SI{}{\celsius}$ & $\SI{}{\newton\per\meter}$ & $\SI{}{\meter\squared\per\second}$ & $\SI{}{\kilo\gram\per\meter\cubed}$ \\
& & ($\SI{}{\centi\meter\cubed\per\second}$) & ($\SI{}{\centi\meter\cubed\per\second}$)&  ($\SI{}{\meter}$)&($\SI{}{\meter}$) \\[0.5ex] 
% & ($\SI{}{\milli\meter}$)&($\SI{}{\milli\meter}$)  % inserts table
%heading

%\hline % inserts single horizontal line

Water & \cite{mohajer2015circular} & 2.5-8.33 & 3058 & 3.83 & 3.92  \\ % inserting body of the table
Water + surfactant & \cite{mohajer2015circular} & 1.66-3.33 & 807.4 & 2.75 & 2.75 \\
Ethylene glycol solution & \cite{bohr1996hydraulic,bohr1997averaging} & 27 & 129.7 & 2.86 & 2.76 \\

 & \cite{rojas2010inertial} &  &  &  &  \\
 Silicone oil - 1  &\cite{duchesne2014constant} & 4.3 &12.44& 2.37 & 2.1\\  
%  Silicon oil - 2  &\cite{duchesne2014constant} & - &2.755& - & -\\

% % & 3.83 & 3.92  \\ % inserting body of the table
% Ethylene glycol   & \cite{saberi2019experimental}& 16.9 - 86.3 &91.6 & - & 2.98\\ 
% & 2.86 & 2.76 \\

% & 2.1 \\
% & 2.36-3.8 \\

% % & 2.1 \\
% % & 2.36-3.8 \\

% % & 2.1 &-  \\
% WP95/5 & \cite{bhagat2018origin} & 83-200 & 42.5&-&2.96   \\
% WP80/20 & \cite{bhagat2018origin} & 83-200  & 26 &-&2.34    \\
% SDBS & \cite{bhagat2018origin} & 83-200 & 38 & -&2.78    \\
%[1ex] % [1ex] adds vertical space
%\hline %inserts single line
\end{tabular}
 % is used to refer this table in the text
\label{table:Table_height}
\end{table}

  %%%%%%%%%%%%%%%%%%%%%%%%%%%%%%%%%%%%%%%%%%%%
\vspace{-3mm}
% {\it The legend doesn't agree with the equation numbers and why not just say solid and dashed curves? I think we should give labels to the different experiments in the table and use these labels on the figures and in the text. Otherwise it is not clear whose data sets are being compared: e.g. which of the 3 silicone oil experiments are shown as the blue line? -m and why such a big correction in this case compared to the other two cases?}
% {I agree with Paul's comment about using labels for the different liquids. They can be introduced in Tables 1 and 2. This will both avoid confusion and save much space associated with references in the text. Also, why use all these subfigure labels? Just put the label (a) and (b) etc. on the plot}

\subsection{Height of the film downstream of the hydraulic jump}
% \cite{hansen1997geometric} {?? \it is this his figure?}{it's from \cite{bohr1996hydraulic}}
Figure \ref{fig:1}(b) compares the predicted jump height $H$ {predicted by (\ref{eq:CAP2})} and radius $R$ {given by (\ref{scaling_relation})} with the experimental measurements reported by \cite{bohr1996hydraulic} {(BO1)}. The theory and experiments show  excellent agreement {for both the jump radius and the height of the film}. The inset figure compares the predicted and measured  {values of} $H$ for different flow rates of water {and surfactant solution} reported by \citep{mohajer2015circular} {(M1, M2)}, and the theory and the experiments again show excellent agreement. \rkb{Notice that reducing surface tension of liquid results in a thinner downstream film which allow both phenomenological model, (\ref{eq:Phenomenological}) and the scaling, (\ref{scaling_relation2}) to be valid and consistent.}   In table \ref{table:Table_height}, we also compare the predicted and measured $H$ values for silicone oil-1 at a low flow rate -- surface tension dominated jump regime when $Q<Q_C^*$ -- {(\cite{duchesne2014constant}, D1)}, and there is {again} good agreement. 

%\cite{duchesne2014constant} conducted experiments with silicone oils 1-3 for a range of flow rates but for other than the lowest flow rate, $Q$ is either comparable or larger than $Q_C^*$, implying gravity does play a significant role, which we will show in the following section. 
% {Does this statement relate to Figure 1(a), Figure 1(b), or Table 2?}
% \vspace{-2mm}

\subsection{Terrestrial Hydraulic jump experiments}
\label{Hydraulic jump radius}
Figure \ref{Hydraulic_jump_fig} compares the measurements of hydraulic jump radius reported by the studies in Table \ref{table:table1}.
%\cite{mohajer2015circular, bohr1993shallow, brechet1999circular, stevens1991local,choo2016influence, saberi2019experimental, duchesne2014constant, hansen1997geometric}, with the theoretical predictions. 
The \cite{saberi2019experimental,duchesne2014constant,hansen1997geometric} 
Experiments {S1, D4, D5, and H2) all feature}  $Q$ values comparable to or larger than $Q_C^*$, while in all the other cases $Q \ll Q_C^*$. 
% {\it Again it would be helpful to have the $Q$ values in the table.} 
{In all cases the {theoretical curve} passes through the data, except for some small {mismatch} at high jet flow rates, showing that the theory including surface tension and gravity predicts the jump radius accurately. 

\rkb{Figure \ref{Hydraulic_jump_fig}(a) compares the jump radius predicted by the \cite{bohr1993shallow} model with their experimental data, and the theory over-predicts the data. Figure \ref{Hydraulic_jump_fig}(b) shows  \cite{brechet1999circular}'s experimental data alongside with the predictions of our theory and \cite{wang2019role}'s. Our theory gives an excellent prediction while in this surface tension dominated case \cite{wang2019role} overpredicts the data.}   }
%%%%%%%%%%%%%%%%%%%%%%%%%%%%%Figure 
\begin{figure}
\centering
\begin{subfigure}[h]{0.45\linewidth}
\includegraphics[width=\linewidth]{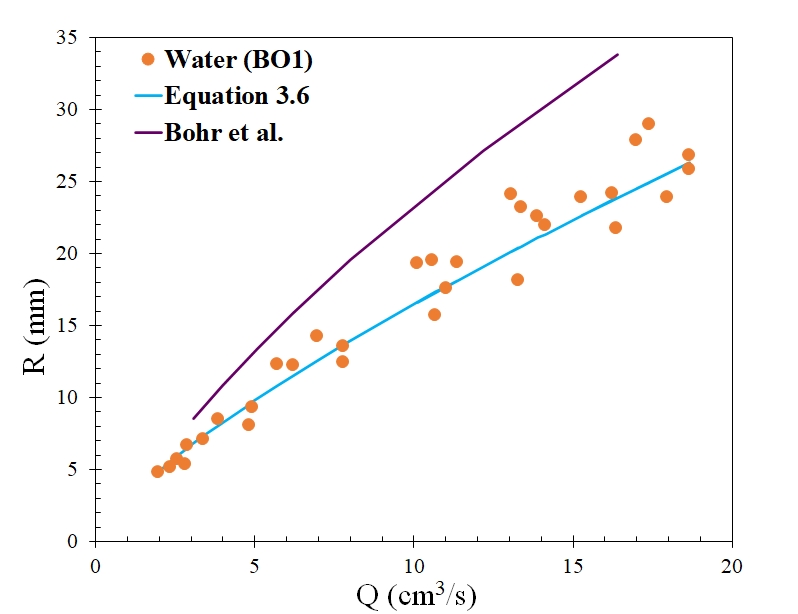}
 \caption{\cite{bohr1993shallow}}
%\label{fig:HJ_pic(b)}
\end{subfigure}%
\hfill
% \vspace{-1mm}
\begin{subfigure}[h]{0.45\linewidth}
\includegraphics[width=\linewidth]{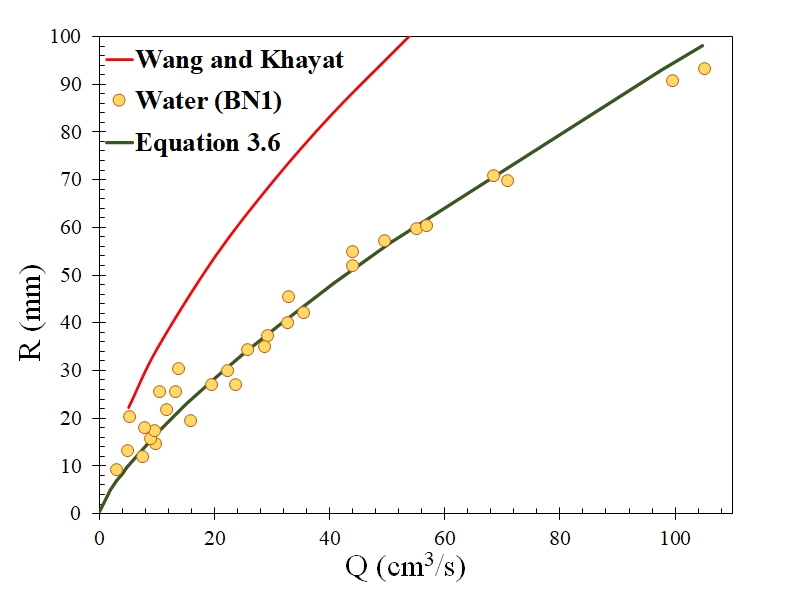}
\caption{\cite{brechet1999circular}}
%\label{fig:HJ_pic_(c)}
\end{subfigure}%
\hfill
\begin{subfigure}[h]{0.45\linewidth}
\includegraphics[width=\linewidth]{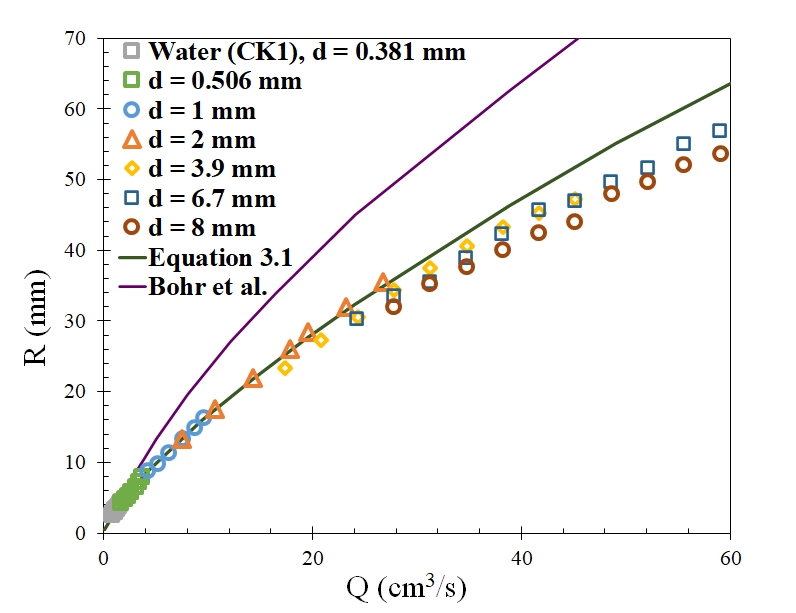}
\caption{ \cite{choo2016influence}}
%\label{fig:HJ_pic_(c)}
\end{subfigure}%
\hfill
\begin{subfigure}[h]{0.45\linewidth}
\includegraphics[width=\linewidth]{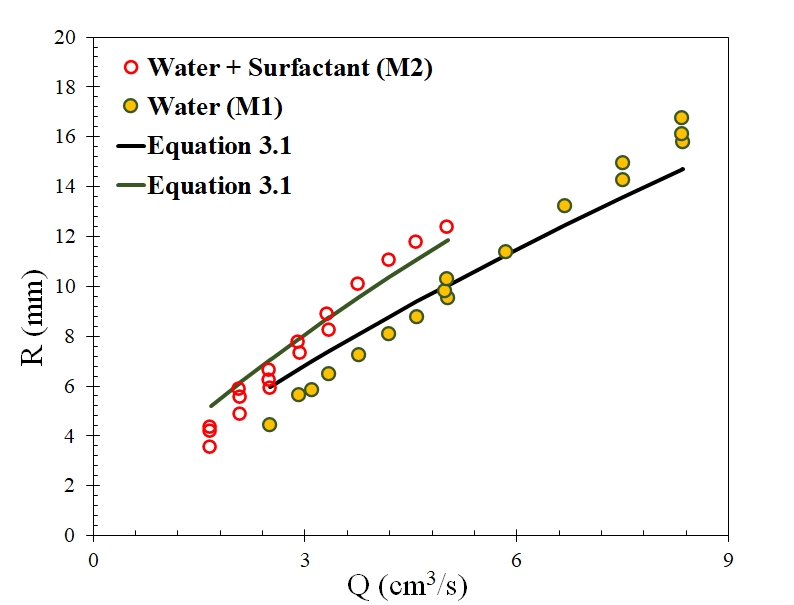}
 \caption{\cite{mohajer2015circular}}
%\label{fig:HJ_pic}
\end{subfigure}
\hfill
% \vspace{-1mm}
% \vspace{-3mm}
\begin{subfigure}[h]{0.45\linewidth}
\includegraphics[width=\linewidth]{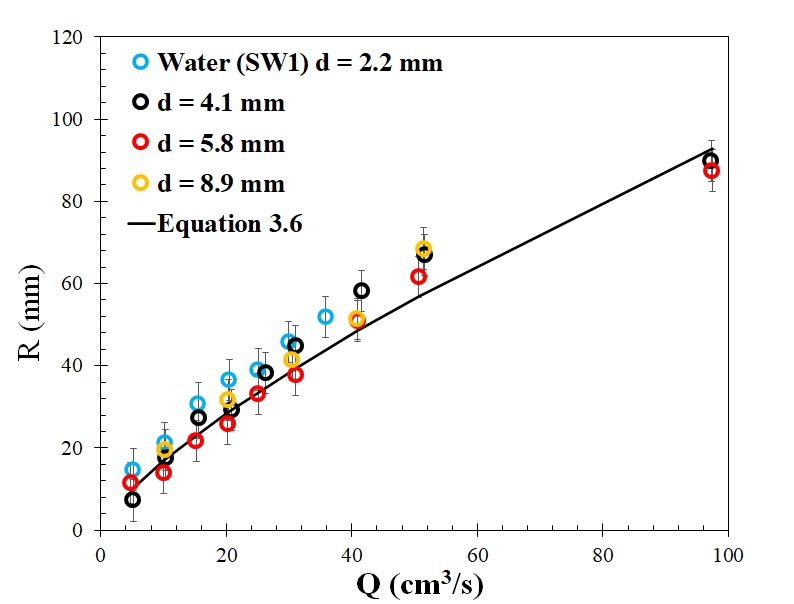}
\caption{\cite{stevens1991local}}
%\label{fig:HJ_pic_(c)}
\end{subfigure}%
\hfill
% \vspace{-3mm}
% \vspace{-3mm}
\begin{subfigure}[h]{0.45\linewidth}
\includegraphics[width=\linewidth]{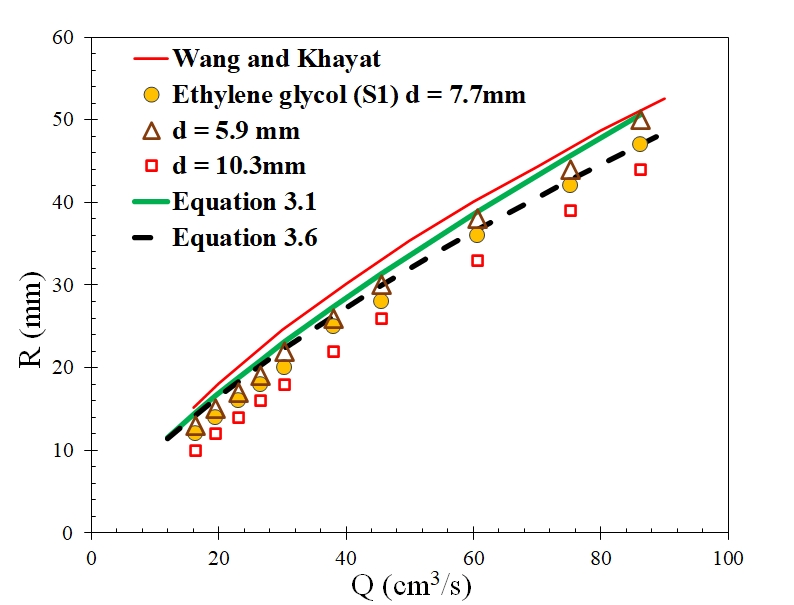}
 \caption{\cite{saberi2019experimental}}
%\label{fig:HJ_pic_(c)}
\end{subfigure}%#
\hfill
% \vspace{5mm}
\begin{subfigure}[h]{0.45\linewidth}
\includegraphics[width=\linewidth]{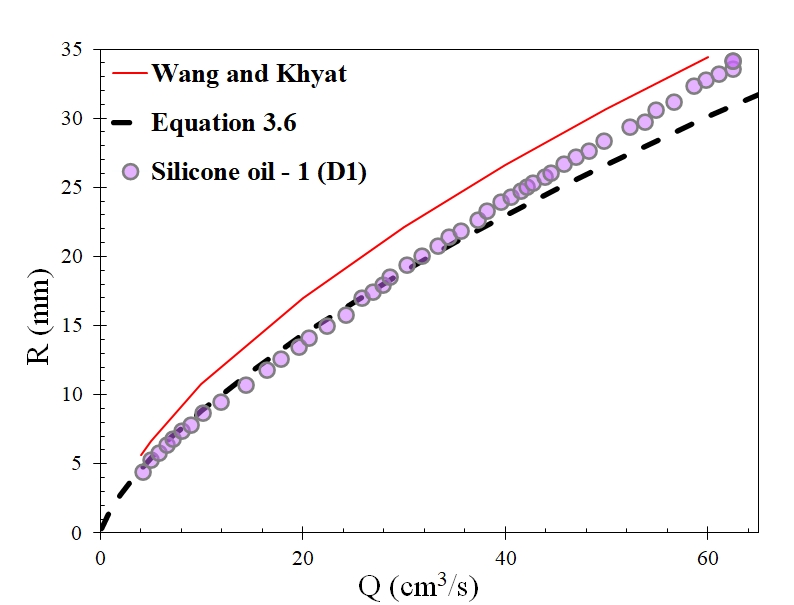}
\caption{ \cite{duchesne2014constant}}
%\label{fig:HJ_pic_(c)}
\end{subfigure}%
\hfill
\begin{subfigure}[h]{0.45\linewidth}
\includegraphics[width=\linewidth]{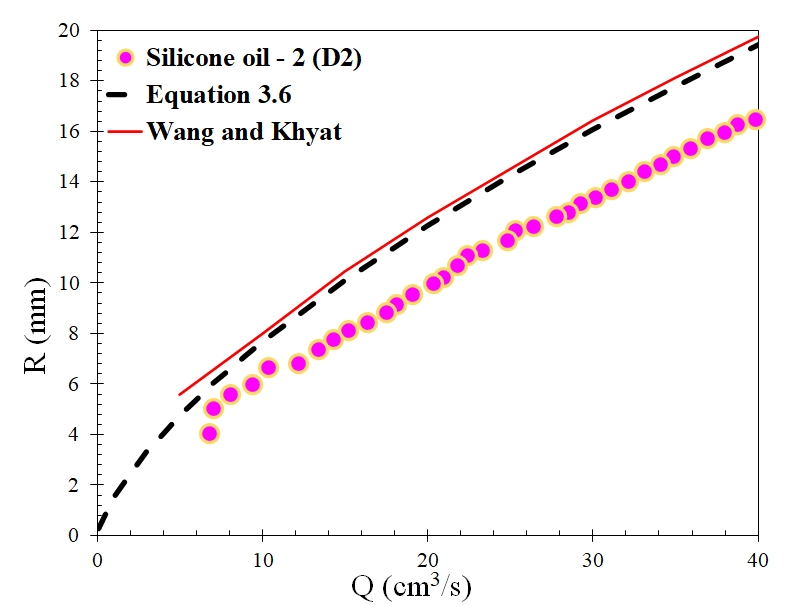}
\caption{ \cite{duchesne2014constant}}
%\label{fig:HJ_pic_(c)}
\end{subfigure}%
\hfill
\end{figure}
\begin{figure}\ContinuedFloat
\begin{subfigure}[h]{0.45\linewidth}
\includegraphics[width=\linewidth]{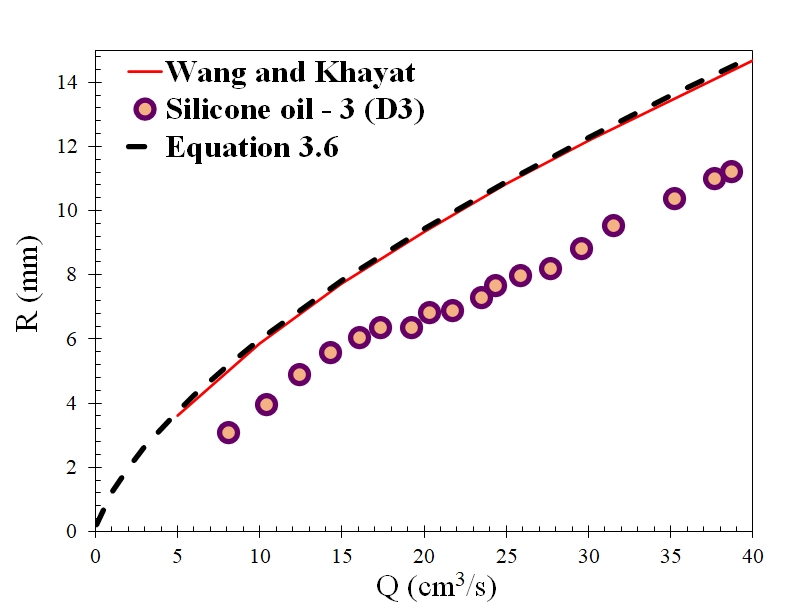}
\caption{ \cite{duchesne2014constant}}
%\label{fig:HJ_pic_(c)}
\end{subfigure}%
\hfill
\begin{subfigure}[h]{0.45\linewidth}
\includegraphics[width=\linewidth]{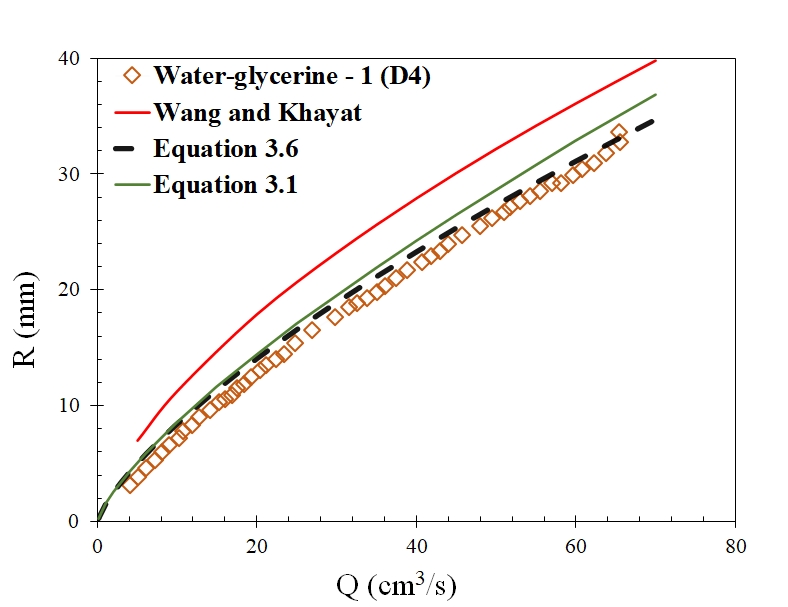}
\caption{ \cite{duchesne2014constant}}
%\label{fig:HJ_pic_(c)}
\end{subfigure}%
\hfill
\begin{subfigure}[h]{0.45\linewidth}
\includegraphics[width=\linewidth]{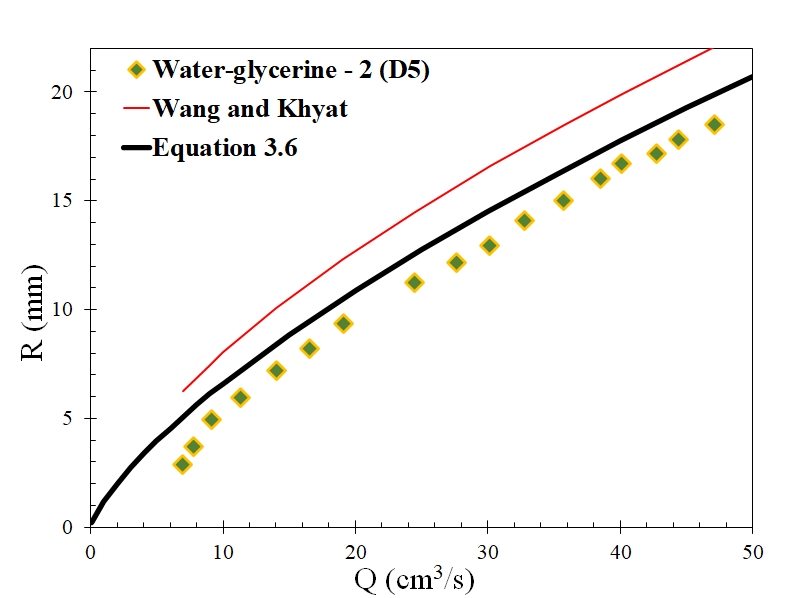}
\caption{ \cite{duchesne2014constant}}
%\label{fig:HJ_pic_(c)}
\end{subfigure}%
\hfill
\begin{subfigure}[h]{0.45\linewidth}
\includegraphics[width=\linewidth]{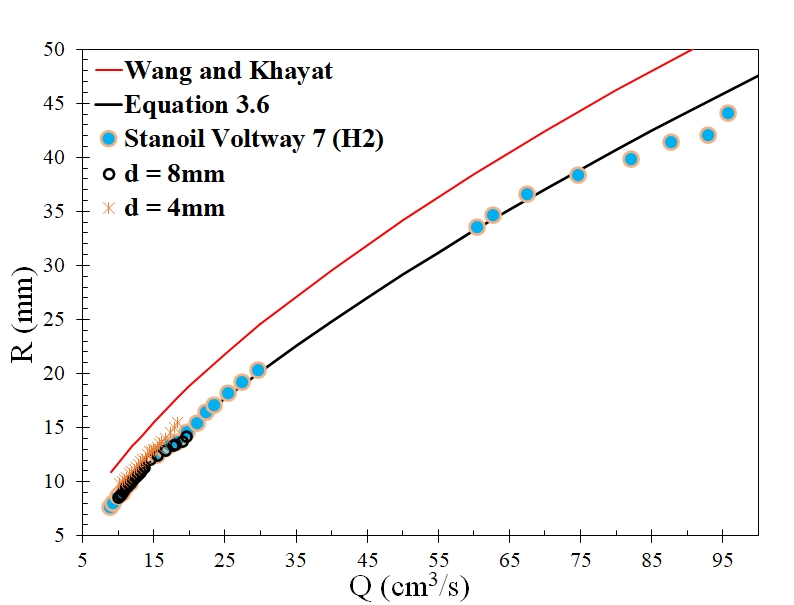}
 \caption{\cite{hansen1997geometric}}
%\label{fig:HJ_pic_(c)}
\end{subfigure}%#

% \vspace{-3mm}

% \setlength{\belowcaptionskip}{-10pt}
% \vspace{5pt}
\caption{Comparison of circular hydraulic jump radii with the theoretical predictions from \citep{bhagat2018origin} for (a) \cite{ bohr1993shallow} water, distance travelled by the jet varied {BO2}, (b) \cite{brechet1999circular} water, distance travelled by the film falling off the plate varied {BN1}, (c) \cite{choo2016influence} water, jet diameter $d$ varied {SW1, CK1}, (d) \cite{mohajer2015circular}  water and surfactant solution {M1, M2,} and (e) \cite{ stevens1991local}  (f) \cite{saberi2019experimental} $d$ varied for ethylene glycol solution, {S1, }(g) \cite{duchesne2014constant} silicone oil - 1,{D1}, (h) \cite{duchesne2014constant} silicone oil - 2,{D2}, (i) \cite{duchesne2014constant} silicone oil - 3, {D3},(j) \cite{duchesne2014constant} water-glycerine, {D4}, (k) \cite{duchesne2014constant} water-glycerine, {D4},(l) (j) \cite{hansen1997geometric}, H2, Statoil Voltway-7 .
%\citeauthor{mohajer2015circular} used water and water surfactant solution of surface tension 72 and 37 $\SI{}{\milli\newton\per\meter}$ and varied the surface tension of the liquid. \citeauthor{ bohr1993shallow} used water as a working fluid and varied the distance between the nozzle and the liquid plate. \citeauthor{ brechet1999circular} used water and varied the height of the falling film. \citeauthor{stevens1991local, choo2016influence} used water and varied the jet diameter. \citeauthor{saberi2019experimental} used ethylene glycol as a working fluid and varied the jet diameter. 
}
\label{Hydraulic_jump_fig}
\end{figure}

% {Get rid of all sub-captions: add labels (a) to (j) to each plot to save space. The sentence in the text introducing data sets can refer to the liquid labels we have asked be added to Tables 1 and 2. The Figure caption does not explain plots (i) and (j). I don't think that the caption for (i), Hansen et al. (1997) is correct.}

%\cite{mohajer2015circular} varied the surface tension of water (by adding surfactant) and observed that for the same flow rate, reducing the surface tension increased the jump radius significantly.  Figure \ref{Hydraulic_jump_fig}(a) compares their experimental results with the values predicted by (\ref{scaling_relation}), and shows excellent agreement for both liquids. \cite{bohr1993shallow} reported the jump radius for water where they varied the distance between the nozzle and the plate. Figure \ref{Hydraulic_jump_fig}(b) shows excellent agreement between the experimental data and (\ref{scaling_relation}). Similarly good agreement is evident in  Figure \ref{Hydraulic_jump_fig}(c) for \cite{brechet1999circular}'s results where the fall height of the draining liquid film at the edge of the plate was varied.

% {\it The following 3 paragraphs repeat and sometimes elaborate on the information about the experiments given in \S3.3. I suggest removing most of this text, adding the extra details to \S3.3 and  then concentrate on describing figure 2 -- not that there is much to say }
\cite{choo2016influence} varied the jet diameter, $d$ from $0.381$ to  $ \SI{8}{\milli\meter}$. Their experimental results in figure \ref{Hydraulic_jump_fig}(c) show little effect of $d$ and are predicted reasonably well by (\ref{scaling_relation}), {although there is a slight over prediction at high $Q$ in this case}. \cite{mohajer2015circular} varied the surface tension of water, and our theory gives excellent prediction  experiments carried out using both water and water-surfactant solution shown in figure \ref{Hydraulic_jump_fig}(d).    
\cite{stevens1991local} studied heat transfer by impinging liquid jets. They also conducted some `\textit{coarse}' measurements of the hydraulic jump radius. Figure \ref{Hydraulic_jump_fig}(e) again shows good agreement despite the larger uncertainty in these measurements. 

All the above cases featured $Q \ll Q_C^*$. \cite{saberi2019experimental} studied the hydraulic jump generated by ethylene glycol on flat and curved surfaces, for which  $Q_C^* = \SI{91.6}{\centi\meter\cubed\per\second}$. They considered flow rates up to $\SI{86.3}{\centi\meter\cubed\per\second}$, so some influence of gravity is expected at the higher flow rates.   Figure \ref{Hydraulic_jump_fig}(f) compares their data with the prediction for $R$ considering surface tension as the dominating force, ( \ref{scaling_relation}), and the prediction considering both gravity and surface tension, (\ref{correction}). Both relationships match the data at low $Q$: at higher values the new result, (\ref{correction}), gives excellent agreement with the experimental data. \rkb{Also shown is \cite{wang2019role}'s prediction, which yields a solution when $Fr = 1$. This model over-predicts the data. Nevertheless, the influence of gravity is evident.  }

\cite{duchesne2014constant} studied hydraulic jumps produced by higher viscosity liquids. 
% In this study, for the three silicone oils, $Q \gtrsim Q_C^*$ whereas for water-glycerine solutions, at higher flow rates $Q$ is comparable to $Q_C^*$. For all these liquids, gravity does play a significant role, particularly for the silicone oils2, and 3, and the hydraulic jumps are expected to be {influenced} by gravity.  
{Figure}~\ref{Hydraulic_jump_fig}(g) compares their data for silicone oils with the prediction for $R$. For silicone oil - 1, our theory including both surface tension and gravity gives an excellent prediction of the data, \rkb{while the \cite{wang2019role} model over-predicts the data. }{ {Figures}~\ref{Hydraulic_jump_fig}(h) and (i) compare theory with experiment for silicone oil - 2 and 3, for which $Q\gg Q_C^*$, and in both cases the predictions are not good. In these gravity-dominated cases our theory is in good agreement with \cite{wang2019role}'s theory (neither gives satisfactory agreement), \rkbb{however, \cite{duchesne2014constant}'s model with a fitting parameter gives a good prediction to their experimental data.} The plots for water-glycerine - 1  solutions in Figure \ref{Hydraulic_jump_fig}(j) show good agreement with our theory, while the \cite{wang2019role} model again over-predicts the data. 
Figure \ref{Hydraulic_jump_fig}(k) presents the results for water-glycerine - 2: with $Q_C^* = 39.2 \SI{}{\centi\meter\cubed\per\second}$,  $8<Q<50 \SI{}{\centi\meter\cubed\per\second}$, the influence of gravity is expected to be modest. Our model overpredicts the data slightly while the \cite{wang2019role}'s model over-predicts noticeably more. } 

% It should be noted that the \cite{duchesne2014constant} experiments with water-glycerine solutions and those reported by \cite{bhagat2018origin} employed different orientations, such that the latter did not require correction due to gravity.

\rkb{\cite{hansen1997geometric} reported $R$ values for water, Statoil Voltway 7, and HV-46 (surface tension and density unknown). Figure ~\ref{Hydraulic_jump_fig}(l) compares their $R$ values measured for Statoil Voltway 7 with (\ref{correction}) and these show good agreement, while the \cite{wang2019role} model again over-predicts the data. } 

% water with those of \cite{brechet1999circular,choo2016influence,bohr1993shallow,mohajer2015circular}, and (\ref{scaling_relation}). For $Q\gtrapprox\SI{15}{\centi\meter\cubed\per\second}$, their values are consistently higher than others in the literature and the prediction. However, the experimental data give excellent agreement with (3.1) for surfactant solution. In their paper they mention the use of `\textit{a little soap}' but this can not be confirmed. Their experimental data for Statoil Voltway 7 are compared with (\ref{correction}) in Figure ~\ref{Hydraulic_jump_fig}(j) and these show good agreement.
% {\cite{wang2019role} compared their results with \cite{hansen1997geometric} experimental data and assumed that \cite{hansen1997geometric} used silicone oils, hence the physical properties for the predictions}
\section{Discussion}
\label{Sec:Discussion}
\subsection{Limitations of the phenomenological model}
\rkb{This approach for identifying the jump location (\cite{belanger1828essai, rayleigh1914theory, watson1964radial}) has been based on a balance between the rate of change of momentum to the thrust of pressure, across the jump. In \S\ref{Previous_studies2}, we noted that \cite{watson1964radial} measured the jump height, and calculated momentum flux on either side of the jump to satisfy the phenomenological condition given by (\ref{eq:Phenomenological1}). In \cite{bhagat2018origin, bhagat2019circular} we showed that formation of a thin film hydraulic jump ($Q<Q_C^*$) does not require a downstream film, or the thrust of gravitational pressure, as it occurs at any surface orientation, \rkbb{as well as we have included data for hydraulic jumps in micro-gravity where the thrust of pressure due to gravity is irreverent.} Theoretically the governing equation breaks down at a finite radius, giving a singularity, which experimentally manifests as a jump. At this point the upstream flow contains the information or the condition for critically, which is $We = 1$, while the governing equation neither requires nor gives any information about the flow beyond the hydraulic jump (see §4.4 of \cite{bhagat2019circular}).  For a particular flow configuration -- vertical impingement of a liquid jet on horizontal surface -- continuity of momentum beyond the jump or the point of singularity, implies liquid flow in a thicker sub-critical film, and allow us to connect up and downstream film determining a critical features, the height of the jump. In \S\ref{scaling_theory}, we showed that the height of the jump, $H$ is a manifestation of the flow field whose magnitude depends on the surface tension of the liquid. Conversely, the jump height, $H \approx \sqrt{\frac{2 \gamma }{\rho g}}$ implies at the location of jump, $We \approx 1$.

In the work by \cite{fernandez2019origin}, numerical solutions to the Navier-Stokes equations were presented for a vertical liquid jet impinging normally onto a horizontal plate. In their solution, they essentially applied the phenomenological condition for the jump which gave them the hydraulic jump radius. We have pointed out earlier that in their Figure 5 they compare their predictions with the experimental results obtained for the opposite scenario, of a liquid jet impinging on to a ceiling, and not surprisingly found good agreement (given that \cite{bhagat2018origin} found that the orientation had little effect). They also reported that, if the hydraulic jump is stationary, its radius would be practically independent of the surface tension, which is not consistent with experimental results such as those reported by \cite{mohajer2015circular}.}

\rkb{In summary, the experimental evidence in conjunction of the theory presented in \S\ref{Momentum balance across the jump on a horizontal surface}, although the phenomenological model is a fundamentally consistent physical relation, it does not unravel the underpinning physics of the phenomenon, namely the \textit{origin} of the hydraulic jump}.

\rkb{\subsection{Limitations of previous predictive models} 

\cite{tani1949water}, following \cite{kurihara1946hydraulic}, proposed that the hydraulic jump arises due to an adverse gravitational pressure gradient which results in separation of the flow from the plane, and consequently found the jump condition to be where the local $Fr = 1$. However, when compared with experiments the model gave a poor prediction the experiment.
 \cite{watson1964radial} showed that, for a thin film, the effect of a gravitational pressure gradient could be neglected. In \S\ref{Jump radius including surface tension and gravity} we showed that gravity is only important beyond a critical flow rate $Q_C^*$ -- which depends upon the surface tension and viscosity of the liquid -- and for water, $Q_C^* = 3058 \SI{}{\centi\meter\cubed\per\second}$, which is well beyond kitchen sink scale hydraulic jumps and the experiments presented by these workers. \rkbb{In the cases where the surface tension is weak ($We \ll 1$) and does not cause the jump; the spreading liquid film slows down due to viscosity/wall friction, causing the liquid film thickness to increase while the momentum reduces, and at a critical radius when the condition of criticality $1/We + 1/Fr^2 =1$ is met, the jump occurs mainly due to gravity.}

More recently, \cite{wang2019role} derived an almost identical governing equation -- the differences being the use of a cubic velocity profile and incorporating the effect of gravity in the boundary layer. They stated that the model was valid for a high viscosity liquid, for which they did not offer any physical reasoning, and compared their model with two different experimental data sets. We have compared their model with eight other different experimental data sets and, like \cite{kurihara1946hydraulic, tani1949water}'s model, it gives poor predictions for many of these, and particularly for aqueous solutions. For fluids with low $Q_C^*$ -- for which the experiments were conducted in the gravity dominated regime -- the over-prediction of their model was reduced. In no case did it provide a better prediction than (\ref{correction}). It is noteworthy that the form of $Q_C^*$, with $\nu$ in the denominator, offers insight into why \cite{wang2019role}'s model would be more likely to describe jumps for high viscosity and low surface tension liquids. Moreover, \cite{fernandez2019origin}'s result, obtained with a high viscosity liquid, proved to be in the surface tension dominated regime, \textit{i.e.} $Q< Q_C^*$, as a result of its high surface tension.

% It is instructive to notice \cite{kurihara1946hydraulic, tani1949water} theory which does not account for the surface tension,  predicts a larger jump radius .  that Silicone oils which are low surface tension and 
}

\section{{Conclusions}}
\label{sec:conc}
This paper provides further experimental evidence for the surface tension origin of the kitchen sink circular hydraulic jump that has been presented by \cite{bhagat2018origin,bhagat2019circular}. {In \cite{bhagat2019circular} by applying conservation of momentum in wall normal directions, we showed that the curvature becomes singular at a finite radius when $We = 1$, and a scaling analysis on radial direction momentum also yielded the jump when $We = 1$. 

% \rkb{We also reviewed previously reported predictive and phenomenological modelling approaches. The predictive modelling approach first started by \cite{kurihara1946hydraulic,tani1949water} and later by \cite{kasimov2008stationary, wang2019role} effectively search for the solution $Fr =1$, where their equation becomes singular, which \cite{kurihara1946hydraulic,tani1949water} have already reported gives a over-prediction of the experimental data. }
}

Including gravity, \cite{bhagat2018origin} found that formation of the hydraulic jump occurred when $1/We + 1/Fr^2 = 1$. For kitchen sink jumps with water,  $Fr^2 \gg 1$, and the jump radius is determined by $We = 1$, {\it i.e} by the surface tension of the liquid alone. We {here} recognise that beyond a critical flow rate, $Q_C^*$, gravity could play a role. For water at room temperature, $Q_C^* =\SI{3058 }{\centi\meter\cubed\per\second}$, which is much larger than typical kitchen sink flow rates. For other liquids, such as silicone oils with low surface tension and high viscosity, gravity \textit{can} be important.  Consequently, we derived a scaling relation, (\ref{correction}), incorporating both surface tension and gravity. \rkb{We also compared experimental data with the recently published model of \cite{wang2019role}, which gives over-predicts the experimental data. For cases where $Q \ll Q_C^*$, their model gives large over-predictions, while for low surface tension and high viscosity liquids, for which $Q \gg Q_C^*$, the over-prediction is smaller.}

\rkb{We revisited the phenomenological models, such as those of \cite{rayleigh1914theory, watson1964radial, bush2003influence}, which are essentially statements of continuity of momentum at the hydraulic jump. If the approximate governing equation lacks the information for the origin of hydraulic jumps -- for example, \cite{watson1964radial}'s theory, which ignores both gravity and surface tension in the governing equation -- the phenomenology will still be valid (continuity of momentum), however the model itself lacks the information about the origin of the hydraulic jump.}

\rkb{In this study, now we have combined the phenomenological and our predictive theory which allowed us to predict a previously un-predicted feature, the height of the jump of the $We \approx 1$ case.}  We applied the radial momentum balance across the jump in combination with the information from our theory that at the jump, $We \approx 1$,  yielding the jump height, $H \approx \sqrt{2\gamma/\rho g} $, in good agreement with the experiments reported in the literature. \rkb{\cite{bush2003influence} studied the influence of surface tension on hydraulic jumps -- they rightly added the hoop stress term due to the interfacial tension, $\gamma \frac{\Delta H}{R}$, in the phenomenological model, which is indeed very small when $\frac{\Delta H}{R}$ is small. They also conducted circular hydraulic jump experiments with a weir to manipulate the height of the jump, and indeed found their conclusion that for kitchen sink scale jump, $\gamma \frac{\Delta H}{R} \to 0$ hence it can be ignored. However, we have now shown that, unless the downstream flow is manipulated, the height of the jump is determined by the jump condition $We \approx 1$. 

% and the phenomenological model is impoverished of information on the origin of the hydraulic jump and does not allow to state about it. 
}
% \diw{I don't think this bit is needed - it is not a new conclusion} Furthermore, our previous work \cite{bhagat2018origin} we showed that irrespective of the orientation of the surface, the jump appears at the same location.\cite{fernandez2019origin} in their numerical study applies the continuity of momentum at the jump, consequently, they found an agreement between their predicted jump and the jump radius measured in opposite orientation, for the scenario when liquid jet was impinged on a ceiling.  }
\rkbb{We present experimental data for the hydraulic jump experiments carried out in micro-gravity. The existence of the hydraulic jump in micro-gravity unequivocally shows that gravity can't be the principal force in the formation of a hydraulic jump. Furthermore, our theory gives a good prediction to the experimental data for the hydraulic jumps in micro-gravity, confirming that kitchen sink scale hydraulic jumps are created due to surface tension.} Subsequently, we compared the theory with experimental measurements in the literature reporting circular hydraulic jump radii for water, aqueous surfactant solution, ethylene glycol, ethylene glycol/water solution, three different silicone oils, water/glycerine solutions and a lubricating oil. The experimental studies encompass wide variations in fluid physical parameters (density, viscosity and surface tension) as well as nozzle diameters, the distance between the nozzle and the plate, and the height of the falling film. The agreement between theory and experiments is excellent. For fluids such as water, surfactant solution and ethylene glycol, for which $Q \ll Q_C^* $ the hydraulic jump radius is solely determined by surface tension, viscosity and the density of the liquid, and is captured by (\ref{scaling_relation}). For high viscosity ethylene glycol solutions and silicone oil - 1, at high flow rates, when $Q\sim Q_C^*$ gravity starts to influence the jump, nevertheless the jump is dominated by surface tension. For high viscosity and low surface tension silicone oils 2 and 3, for which $Q\gtrapprox Q_C^*$, gravity and surface tension both play a role. 

This work, in combination with \cite{bhagat2018origin, bhagat2019circular} provides a coherent account of the origin of hydraulic jumps. It also demonstrates the role, relevance and influence of surface tension in interfacial flows. 
%{In \cite{bhagat2018origin} we have shown that the current consensus that influence of surface tension in hydrodynamics is contained in Laplace pressure is flawed and we have given the right boundary condition for the interfacial flow.
\vspace{-7mm}
\section*{Acknowledgements}
\vspace{-1mm}
We thank Profs. Kim and Webb for sharing data, Prof. Dimon for the information regarding the fluids used for the experiment in \cite{hansen1997geometric}, and Prof. Li and Dr Mohajer for providing the liquid film thickness measurements for their surfactant solutions. \rkbb{ We would also like to thank Prof John Kuhlman, Dr Mehran Mohebbi, Mr Kyle G. Phillips, Mr Tristan Wolfe, and Dr Kerri Phillips for providing the raw data and videos of the micro-gravity experiments.}     \\

\textbf{Declaration of Interests:} The authors report no conflicts of interest.

\bibliographystyle{jfm}

\bibliography{jfm-instructions.bbl}

\end{document}